\pdfoutput=1

\documentclass{SciPost}

\hypersetup{
    colorlinks,
    linkcolor={red!50!black},
    citecolor={blue!50!black},
    urlcolor={blue!80!black}
}

\makeatletter
\let\scipost@latex@note\@latex@note
\def\@latex@note#1{}
\makeatother
\usepackage[bitstream-charter]{mathdesign}
\makeatletter
\let\@latex@note\scipost@latex@note
\makeatother
\usepackage{graphicx}
\usepackage{comment}
\usepackage{braket}
\usepackage[normalem]{ulem}
\usepackage{amsmath}
\DeclareSymbolFont{usualmathcal}{OMS}{cmsy}{m}{n}
\DeclareSymbolFontAlphabet{\mathcal}{usualmathcal}

\newcommand{\bluehead}[1]{\texorpdfstring{\color{scipostdeepblue}{#1}}{#1}}

\fancypagestyle{SPstyle}{
\fancyhf{}
\lhead{\colorbox{scipostblue}{\bf \color{white} ~SciPost Physics }}
\rhead{{\bf \color{scipostdeepblue} ~Submission }}

\fancyfoot[C]{\textbf{\thepage}}
}
\usepackage{xcolor}

\begin{document}

\pagestyle{SPstyle}

\begin{center}{\Large \textbf{\color{scipostdeepblue}{
Unpaired fractional excitations from many-body quantum geometry
}}}\end{center}

\begin{center}\textbf{
O. Kashuba\textsuperscript{1$\dagger$},
Ram Mummadavarapu\textsuperscript{1$\dagger\star$} and
R.-P. Riwar\textsuperscript{1}
}\end{center}
\begin{center}
{\bf 1} Peter Gr\"unberg Institute, Theoretical Nanoelectronics, Forschungszentrum J\"ulich, D-52425 J\"ulich, Germany
\\[\baselineskip]
$\star$ \href{mailto:r.mummadavarapu@fz-juelich.de}{\small r.mummadavarapu@fz-juelich.de}
\\[\baselineskip]
$\dagger$ {\small These authors contributed equally to this work.}
\end{center}

\section*{\bluehead{Abstract}}

Compact scalar fields on lattices describe a broad class of many-body systems, from superconducting circuits to interacting bosons and spin systems. We show that a generic quantum geometric coupling between such fields induces a quantized Chern response, which in superconducting circuits can be interpreted as an integer-quantized gyration conductance. Because the fields are compact, these couplings produce Landau levels with finite degeneracy. Projected onto the lowest level, Josephson operators acting on different nodes acquire fractional exchange phases set by the Chern numbers. As a consequence, a junction shunting one node becomes periodic in a fraction of the flux quantum $h/2e$, and a gate charge on another node produces a charge counterpart of the fractional Josephson effect. In suitable networks, these operators encode Majorana fermions and their parafermion generalizations, each switched on by a single local Josephson junction, while the network itself supplies the nonlocal links that otherwise require long gate sequences. Single junctions then generate Hamiltonian terms containing unpaired encoded excitations, including parity-odd terms forbidden by Wigner superselection for elementary fermions, suggesting a route to all-to-all gates in fermionic quantum simulators.
\section{\bluehead{Introduction}}

\begin{figure}
\centering
\includegraphics[width=0.5\textwidth]{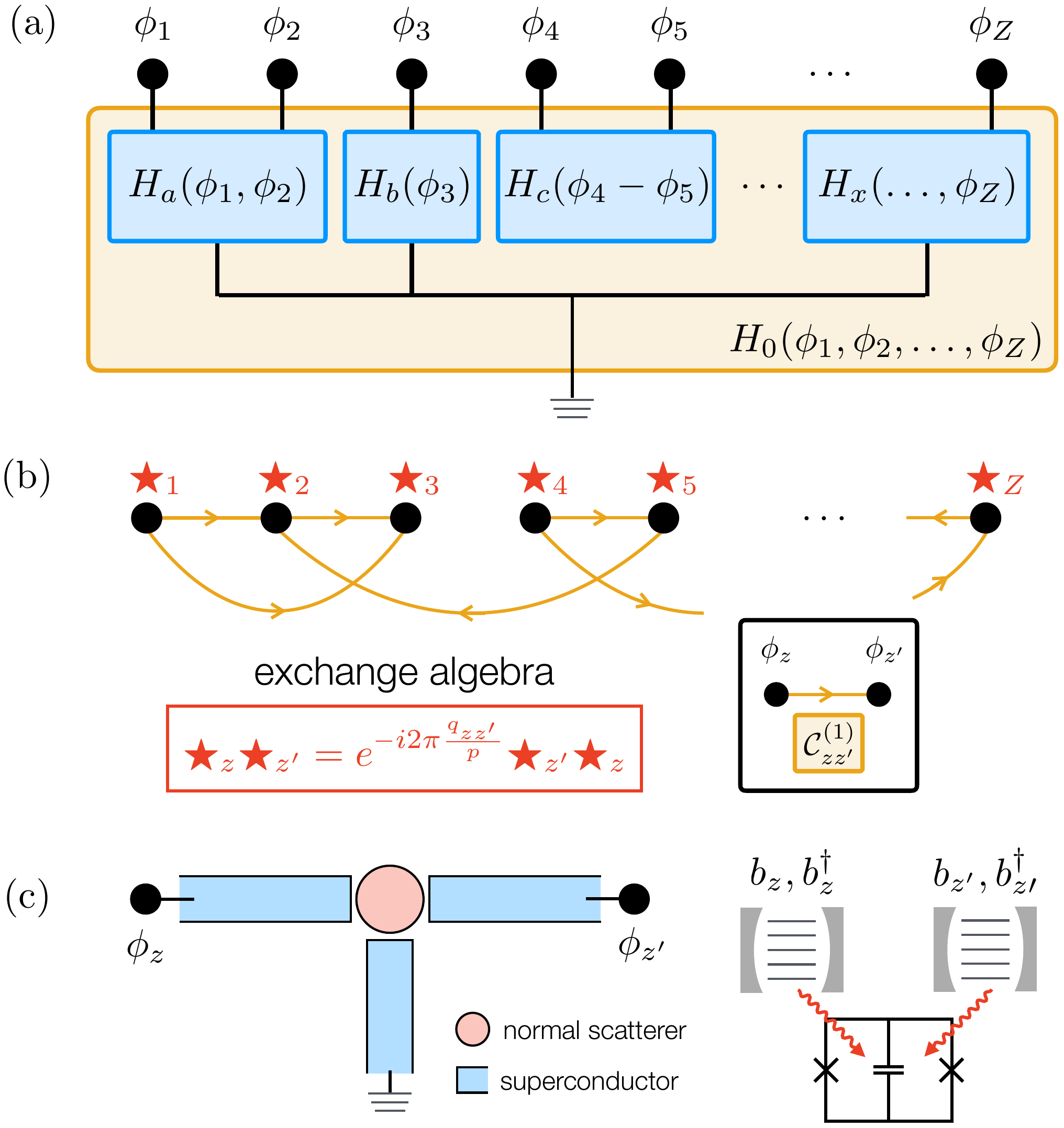}
\caption{Main result of our work. (a) We consider a lattice with nodes $z$ described by compact fields $\phi_z$, where generic couplings are mediated by extra degrees of freedom described by Hamiltonians $H_{a,b,\ldots}$, summarized into a total multi-terminal Hamiltonian $H_0$. (b) By including many-body quantum fluctuations, the quantum geometric contribution of $H_0$ maps the lattice onto a network of circuit nodes coupled by gyrators with quantized gyration conductance given by a (first) Chern number. After projection, each local phase harmonic acts as a magnetic-translation operator, with rational commutation phases set by the Chern matrix. (c) Possible realizations of coupler systems between two nodes with nonzero Chern number, either with superconducting circuits (left) or quantum cavities (right).}
\label{fig:qcgyrator}
\end{figure}

Networks of compact phase variables arise in superconducting circuits such as Josephson-junction arrays~\cite{Nazarov_1995,Fazio_2001}, as well as in bosonic platforms such as coupled optical cavities and ultracold atomic gases. The phases can be coupled through devices with gapped internal degrees of freedom, including multiterminal Josephson junctions whose many-body ground state depends on the phases~\cite{Riwar2016}. Eliminating these internal degrees of freedom in a geometric Born--Oppenheimer approximation~\cite{Mead_1979} produces a Berry curvature that acts as an effective magnetic field for the phases~\cite{riwar_2023discrete,Matute_2024}. In superconducting circuits, this geometric response provides a gyrator coupling for slow phase dynamics at energies well below the internal excitation gap, see Fig.~\ref{fig:qcgyrator}. Its microscopic origin distinguishes it from the finite-frequency Hall effect response considered in Ref.~\cite{Viola_2014} and from the ideal gyrator taken as an input to the circuit model of Ref.~\cite{Rymarz_2021}. Furthermore, compactness of the phase coordinates quantizes the integrated Berry curvature, fixing the topological contribution to the gyration conductance in units of $(2e)^2/h$.

Our work is motivated by an approach~\cite{Viola_2014,Riwar2016,Rymarz_2021} that recently emerged in the context of quantum circuits, aimed at reducing hardware requirements. Instead of defining topological properties in real space, they can be defined in the abstract space spanned by the compact scalar fields $\phi_z$, where $z=1,\ldots,Z$ labels the nodes of the network and $Z$ is the total number of nodes. These fields describe, for example, superconducting phases $\Delta e^{i\phi_z}$ or bosonic phases $b_z^{(\dagger)}\sim e^{\pm i\phi_z}$. In this context, non-reciprocal gyrator elements have recently gained significant traction~\cite{Viola_2014,parra2019canonical}. They have been proposed~\cite{Rymarz_2021} to realize an effective magnetic field on a 2D lattice in $\phi$-space, realizing the Hofstadter butterfly spectrum at non-integer gyration conductance.

The compactness of the field $\phi$ plays a pivotal role. For superconducting circuits, this compactness represents charge quantization, whose importance has been widely studied~\cite{Yurke_1984,Loss_1991,Koliofoti_2023}\footnote{Whether or not the superconducting phase $\phi$ is compact (or more precisely, closed) depends on the existence of a mechanism to entangle with the number turns a given system performs in $\phi$-space; see Refs.~\cite{Loss_1991,Mullen_1993,Koliofoti_2023}.}. A compact $\phi$ space provides a natural analogue to the Brillouin zone, forming a higher dimensional torus $\mathbb{T}^Z$. Any gapped eigenspectrum on that Brillouin zone has associated with it a Berry curvature, whose integral yields a quantized Chern number.
For multi-terminal Josephson junctions, this generic feature has been predicted to allow for the creation of topological matter in higher dimensions~\cite{Riwar2016} without necessarily involving exotic materials, simply by adding a sufficiently large number of circuit nodes. This mechanism has been explored in various contexts, including Andreev bound states~\cite{Riwar2016,Eriksson2017,Meyer:2017aa}, tunnel junctions~\cite{Fatemi_2021,Peyruchat_2021,Herrig_2022}, and quantum dots~\cite{klees2021ground,teshler2023ground}, as well as generalizations to higher-order invariants~\cite{Weisbrich_2021}, and has spurred considerable progress on the experimental side in the control and measurement of multiterminal Josephson junctions~\cite{Strambini:2016aa,Pankratova_2020_multitJJexp,Coraiola_2023,Peyruchat_2024}. Recently, there have emerged some early efforts to explore the impact of the Berry curvature on qubit architectures~\cite{riwar_2023discrete,Matute_2024,Virtanen_2024}. 
These works analysed first consequences for small circuits. The quantum problem of an arbitrary network of phases coupled in this way, and the nature of the operators that govern its low-energy physics, have remained open.

We solve this problem for an arbitrary number of nodes. Mathematically, it is a compact, multidimensional version of the ``topological'' (Chern--Simons) quantum mechanics of Dunne, Jackiw and Trugenberger~\cite{Dunne_1990}, with a matrix of integer Chern numbers in place of a single magnetic field. The central relation of this work is the following. After projection to a fixed Landau level of the phase dynamics, constructed in Sec.~\ref{sec_LL_torus}, the compact-phase operators $\eta_z\equiv e^{i\Phi_z}$ satisfy
\begin{equation}\label{eq:chern_circuit_exchange}
    \eta_z\eta_{z'}
    =
    e^{-i2\pi q_{zz'}/p}
    \eta_{z'}\eta_z.
\end{equation}
The integer first Chern numbers associated with pairs of phase coordinates form a Chern matrix, whose inverse determines the pair-dependent exchange phases $q_{zz'}/p$, depending explicitly on the node labels $z,z^\prime$. Throughout this work, we refer to the operators $\eta_z$ satisfying the exchange algebra in Eq.~\eqref{eq:chern_circuit_exchange} as \emph{projected phase operators}. These operators act within the degenerate subspace of a Landau level of a single collective coordinate on the phase torus.

Fractional exchange phases, such as those in Eq.~\eqref{eq:chern_circuit_exchange}, are familiar from the fractional quantum Hall effect, where they accompany fractionally charged quasiparticles~\cite{Wen_1990,Tong_2016}. Here, however, charge is quantized in integer units from the outset, while transferring a single Cooper pair induces a fractional flux. Integer charge quantization permits interaction terms of the form $\eta_z^m\eta_{z'}^{m'}$, with $m,m'\in\mathbb Z$, without requiring conservation of the total node charge, since Cooper pairs can be exchanged with ground. In contrast to uniform chiral-edge and parafermion algebras, where the exchange relation is controlled by a single statistical angle, the present construction yields a pair-dependent exchange matrix and therefore a pair-dependent family of projected charge-conserving operator products~\cite{Wen_1990,Tong_2016,Fendley_2012}. Moreover, upon including a trivial node $z_0$, for which $\eta_{z_0}=\mathbf{1}_{z_0}$, the effective low-energy Hamiltonian can contain terms of the form $\mathbf{1}_{z_0}\eta_z+\mathrm{h.c.}$, corresponding to an unpaired projected operator and thereby breaking the Wigner superselection rule (in the non-relativistic setting considered here).
This allows communication protocols between nodes that use only local controls, which links the present construction both to quantum hardware applications and to broader foundational questions about locality and superselection in quantum mechanics~\cite{Peres_2004,Bub_2005,Chiribella_2011}, specifically to how local realism and Wigner superselection for fermionic systems~\cite{Friis_2016,Vidal_2022} may be formulated on an axiomatic level.

Charge and flux control, combined with tunable Josephson junctions, provide gates that act directly on the projected operators. This could enable simulations of Majorana fermions for quantum chemistry~\cite{McArdle_2020}, solid-state systems~\cite{Barends_2015,Khodaeva_2024} and the Sachdev--Ye--Kitaev model~\cite{Sachdev_1993,Kitaev_2015} with local control. In a Jordan--Wigner encoding, fermionic operators can map to strings of qubit operators~\cite{Whitfield_2011}. In the Chern circuit, a single junction realizes the corresponding projected operator, and Eq.~\eqref{eq:all_to_all} constructs an encoded two-body gate from three single-node pulses. The projected Landau level also provides a finite magnetic-translation representation of the kind used in Gottesman--Kitaev--Preskill encodings~\cite{Gottesman_2001,Rymarz_2021}.

This work is structured as follows. We first review the relevant properties of compact fields and then formulate the conditions under which the projected phase-operator algebra emerges (Secs.~\ref{sec_nonreciprocity}--\ref{sec_LL_torus}). Specifically, we assume a coupler whose excitation gap lies well above, and whose ground-band variation lies well below, the cyclotron energy. We consider an even number of nodes with a nonsingular Chern matrix and Josephson couplings too weak to mix Landau levels. Under these assumptions, we establish the following main results.
\begin{enumerate}
\item The topological contribution to the coupler's geometric response gives a gyration conductance $G=(2e)^2\mathcal C^{(1)}/h$, Eq.~\eqref{eq_g_matrix}. A nonsingular Chern matrix requires the coupler to exchange Cooper pairs with ground, Eq.~\eqref{eq:ground_lemma} (Secs.~\ref{sec_gyration_conductance_quantization}--\ref{sec_LL_torus}).

\item For the ideal Landau Hamiltonian, each Landau level on the phase torus is $|\mathrm{Pf}\,\mathcal C^{(1)}|$-fold degenerate and carries a single irreducible representation of a product of clock--shift pairs, Eq.~\eqref{eq:frobenius} (Sec.~\ref{sec_LL_torus}).

\item A junction shunting node $z$ has flux period $\Phi_0/g_z$, with $g_z$ the greatest common divisor of the Chern numbers at node $z$. An offset charge on another node produces a fractional dual Josephson effect, Eq.~\eqref{eq:dual_majorana} (Sec.~\ref{sec_charge_flux}).

\item For a complete graph with $\mathcal C^{(1)}_{zz'}=2$ for $z<z'$, each Majorana operator is the projection of a single Josephson term, Eq.~\eqref{eq_gamma_z} (Sec.~\ref{sec_simulation_fermion}).
\end{enumerate}
Sections~\ref{sec_fermion_parity} and~\ref{sec_nonidentical} then examine the parity and locality properties of the encoded algebras. We conclude in Sec.~\ref{sec_conclusions_outlook} with a discussion of the relation to chiral-edge theories of the fractional quantum Hall effect.

\section{\bluehead{Preface: compact scalar fields}}

To set the stage, we consider a network of $Z$ lattice points (from now on called nodes). To each node $z$ ($z=1,\ldots,Z$), we assign a field $\phi_z$, which we refer to as a phase. Quantum theories for such a field can be generically expressed via a Lagrangian, $L(\{\dot{\phi}_z,\phi_z\})$, depending on all phases and their phase velocities. The Hamiltonian is obtained from the Lagrangian by the usual Legendre transformation within ordinary canonical quantization, $H=\sum_z n_z \dot{\phi}_z-L$, where the canonical momentum $n_z=\partial_{\dot{\phi_z}}L$ satisfies the regular commutation relation $[\phi_z,n_{z^\prime}]=i\delta_{zz^\prime}$.

Let $\phi_z$ be compact (on a ring of size $2\pi$). The configuration space is thus a $Z$-dimensional torus, and the overall wave function $\psi(\{\phi_z\})$ must be single-valued on that torus. Each node is thus a quantum rotor, and the network is a set of coupled quantum rotors.
Such scalar compact field theories describe a considerable number of systems. For superconducting quantum circuit networks, $\phi_z$ corresponds to the superconducting phase, and $n_z$ is the Cooper pair number on a given node. A bosonic system (e.g., quantum optics) with ladder operators $b_z^{\dagger}$ can likewise be expressed with such fields, most conveniently in the limit of dense bosons~\cite{Bruder_2005}, where the ladder operators are approximated as $b_z^\dagger \sim e^{i\phi_z}$ ($b_z \sim e^{-i\phi_z}$) and the boson number is $b^\dagger_z b_z\sim n_z$. Compact scalar fields also describe chiral Luttinger liquids~\cite{Tong_2016}, the XXZ model~\cite{10.21468/SciPostPhys.15.2.051} and, via the Villain transformation, XY, Coulomb-gas and solid-on-solid models~\cite{Chui_1976}. In most of these cases, $\phi_z$ describes a collective \textit{many-body} degree of freedom.

The Lagrangian typically has the following generic form,
\begin{equation}
    L_0=\frac{1}{8e^2}\sum_{zz^\prime}\dot{\phi}_zC_{zz^\prime}\dot{\phi}_{z^\prime}-\epsilon_0(\{\phi_z\})\ ,
\end{equation}
where the kinetic term proportional to the phase velocities provides quantum fluctuations of $\phi_z$ (once the theory is quantized). Here, $\epsilon_0=\epsilon_0(\{\phi_z\})$ is a potential energy depending only on the phase coordinates and not on their velocities $\dot{\phi}_z$.
Compactness of the phase here can be guaranteed by imposing that the potential energy $\epsilon_0$ be single-valued on the $Z$-dimensional torus -- or in other words, $2\pi$-periodic in all $\phi_z$.

The central physics that we focus on here comes in the form of an \textit{additional} nonreciprocal term, $L=L_0+L_G$,
\begin{equation}\label{eq_L_gyrator}
    L_G=-\sum_z\dot{\phi}_z A_z\ ,
\end{equation}
where the vector-potential $A_z$ is likewise a function of all phases. One might naively expect that $A_z$ likewise needs to be single-valued on the torus, but in actuality, the theory allows for topological terms linear in $\phi_z$, related to the well-known Dirac monopole quantization argument for quantum mechanics on closed manifolds (see further below for an in-depth discussion).

The first-order term $L_G$ has the structure of the Floreanini--Jackiw action of chiral edge theories~\cite{Zee_1995,Tong_2016}, in which a Chern--Simons level multiplies $\dot\phi\,\partial_x\phi$. There, the term follows from a continuum topological field theory with a uniform level. Here, it is derived from a microscopic coupler Hamiltonian on a lattice of nodes, with an integer Chern number for each pair of nodes that need not be uniform; we do not assume equivalence to a local bulk topological field theory.
Together with the capacitive energy, $L_G$ describes the analogue of a massive charged particle in a uniform magnetic field, which Dunne, Jackiw and Trugenberger introduced as a $(0+1)$-dimensional model of topologically massive gauge theory, ``topological'' (Chern--Simons) quantum mechanics~\cite{Dunne_1990}. In their limit of vanishing mass, only the first-order term survives, the coordinates no longer commute, and the dynamics is confined to the lowest Landau level. Here this limit corresponds to a cyclotron frequency large compared with the Josephson energies, which is the regime studied below. Two differences matter. Their particle moves in the plane, with infinitely degenerate Landau levels; here the configuration space is the torus $\mathbb T^Z$, so the degeneracy is finite. And their field is a single number, whereas here it is a matrix of integer Chern numbers fixed by the coupler.

 Motivated by recent progress within the superconducting circuit community (where the emergence and functioning of non-reciprocal terms similar to those in $L_G$ has been discussed in various contexts~\cite{Viola_2014,Rymarz_2021,riwar_2023discrete,Matute_2024}) we provide in what follows a platform-independent microscopic derivation of topological non-reciprocal interactions via many-body quantum geometry.

\section{\bluehead{Nonreciprocity from quantum geometry}}\label{sec_nonreciprocity}

Our starting point is the generic lattice introduced above, with nodes $\phi_z$. While (as emphasized already) this lattice may describe a wide range of physical systems, for the sake of concreteness we use circuit language in what follows and refer to $\phi_z$ as a phase and to the canonically conjugate $n_z$ as a charge. It is furthermore useful to introduce a reference point for the phases, i.e., all values of $\phi_z$ are to be taken with respect to a phase ground set to zero. Changing this reference point corresponds to a global $U(1)$ shift. Invariance under that shift represents charge conservation. In circuits, this reference point is the actual electrical ground.

We now include a generic coupling between all $Z$ nodes, and (optionally) to ground. While different nodes are in general coupled with different physical connections, we can cast all of these details into a single, all-encompassing ``black box'' circuit element. Under a concise set of assumptions, we will show that the black box generically reduces to a network of pairwise connections each described by a (first) Chern number. While we here (for the sake of generality) consider a generic multiterminal setup where we impose no upper bound on the number of nodes $Z$, we will show in a moment that any arbitrarily large network can always be decomposed into elementary building blocks with an individual nonzero Chern number between two active nodes, see, e.g., circuit diagrams like in Fig.~\ref{fig:qcgyrator}.

We emphasize that the internal degrees of freedom of the black box element are \textit{additional}, intrinsic physical degrees of freedom beyond the scalar field. For superconducting circuits, we think of the black box as a multiterminal Josephson junction (or a network thereof) connecting different islands. In the quantum optics context, we can think of it as a coupler device between cavities. We describe the black box by a Hamiltonian as a function of the phases, $H_0(\{\phi_z\})$. For fixed phases, $H_0(\{\phi_z\})$ acts on these internal degrees of freedom; the phases enter only as parameters, and \(H_0\) is independent of $n_z$ and $\dot{\phi}_z$. The assumed gap therefore refers to the spectrum of the internal coupler. The only further restrictions we place on this system are that it has a nondegenerate ground state with a finite gap to the first excited state, and the above compactness condition, i.e., the Hamiltonian shall be $2\pi$-periodic in all phases $\phi_z$, indicating integer quantization of $n_z$. Finally, the emergence of nonzero Chern numbers will imply time-reversal symmetry breaking.

We now include a kinetic coupling. For circuits, this corresponds to capacitive couplings between circuit nodes to ground and between each other. For quantum optics, the same mechanism accounts for self- and cross-Kerr terms of the cavities. This yields the Hamiltonian
\begin{equation}\label{eq_including_capacitances}
    H=2e^2\sum_{zz^\prime}n_z(C^{-1})_{zz^\prime} n_{z^\prime}+H_0(\{\phi_z\})\ .
\end{equation}
Expressing the black box Hamiltonian in terms of its eigenstates $H_0=\sum_k \epsilon_k\vert k\rangle\langle k\vert$ (the $\phi_z$-dependence of energy $\epsilon_k$ and state $\vert k\rangle$ is no longer explicitly denoted), and fixing the index of the nondegenerate ground state as $k=0$, we perform a geometric version of a Born-Oppenheimer approximation, which was first developed in the context of atomic physics by Mead and Truhlar~\cite{Mead_1979} (and whose importance was recently appreciated for Josephson junctions~\cite{riwar_2023discrete,Matute_2024}), where projecting onto $\vert 0 \rangle$ yields the low-energy description,
\begin{equation}\label{eq_H_low}
    H\approx 2e^2\sum_{zz^\prime}(n_z+A_z)(C^{-1})_{zz^\prime}(n_{z^\prime}+A_{z^\prime})+\epsilon_0\ ,
\end{equation}
with the Berry connections $A_z=-i\langle 0\vert\partial_{\phi_z}\vert 0\rangle$. The Born--Oppenheimer projection is controlled by the gap between the ground state and the excited states of the internal coupler, which we define as
\begin{equation}
    \Delta_{\min}
    =
    \min_{\{\phi_z\}}
    \left[
        \epsilon_1(\{\phi_z\})
        -
        \epsilon_0(\{\phi_z\})
    \right].
    \label{eq:minimum_coupler_gap}
\end{equation}
The projection requires this gap to be large compared with all energy scales associated with the slow phase dynamics. In particular, $\omega_{\max}/\Delta_{\min}\ll1$, where $\omega_{\max}$ denotes the largest cyclotron-level spacing. In addition to the canonical charge $n_z$, there is the kinetic charge $n_z+A_z$, whose commutator (between nodes $z$ and $z^\prime$) is gauge-invariant,
\begin{equation}\label{eq_Nz_commutator}
    [n_z+A_z,n_{z^\prime}+A_{z^\prime}]=i\left(\partial_{\phi_{z^\prime}}A_z-\partial_{\phi_z}A_{z^\prime}\right)=i\mathcal{B}_{zz^\prime}\ ,
\end{equation}
and given by the Berry curvature $\mathcal{B}_{zz^\prime}$.
Note that the kinetic charge is the physical charge on the nodes. For circuits, it determines the electric field between nodes.

A further relevant energy scale is the ground-state bandwidth of the coupler,
\begin{equation}
    \Delta\epsilon_0 = \max_{\{\phi_z\}}\epsilon_0(\{\phi_z\}) - \min_{\{\phi_z\}}\epsilon_0(\{\phi_z\}).
\end{equation}
For the two-node case, with $\omega$ the cyclotron frequency of Sec.~\ref{sec_LL_torus}, the regime assumed throughout this work is therefore $\Delta\epsilon_0\ll\omega\ll\Delta_{\min}$. More generally, the nonlinear energy scales must be smaller than the smallest cyclotron spacing, while the largest cyclotron spacing must remain smaller than $\Delta_{\min}$.

The condition $\Delta\epsilon_0\ll\omega$ is the demanding one, and it constrains the coupler independently of its topology. For a two-level coupler, $H_0=\mathbf v(\{\phi_z\})\cdot\boldsymbol\sigma$, the ground energy is $\epsilon_0=-|\mathbf v|$ and the gap is $2|\mathbf v|$, whereas the Chern numbers depend only on the direction $\mathbf v/|\mathbf v|$. Hence
\begin{equation}\label{eq:flatband}
    \frac{\Delta\epsilon_0}{\Delta_{\min}}=\frac{\max|\mathbf v|-\min|\mathbf v|}{2\min|\mathbf v|}\ ,
\end{equation}
and the hierarchy requires a coupler whose eigenvector winds while its band stays flat on the scale of $\omega$. Flattening $|\mathbf v|$ leaves $\mathcal C^{(1)}$ unchanged, so this is an engineering requirement, not a topological one. For Andreev couplers with transparent channels, $\Delta\epsilon_0$ is of the order of the superconducting gap, and $\Delta\epsilon_0\ll\omega$ fails. The regime considered here therefore requires a coupler engineered for a flat ground band, or compensation of the low harmonics of $\epsilon_0$ by additional junctions. The model of Appendix~\ref{app:cavity_supp} demonstrates a Chern-nontrivial coupler, but it has $\Delta\epsilon_0\geq\Delta_{\min}$ throughout its topological phase.

We thus arrive via a well-defined low-energy approximation on the Hamiltonian level at the very same vector potential term given in the Lagrangian term of Eq.~\eqref{eq_L_gyrator}. Adding charge energy terms and the potential energy $\epsilon_0$, $L=\frac{1}{8e^2}\sum_{zz^\prime} \dot{\phi}_z C_{zz^\prime}\dot{\phi}_{z^\prime}+L_G-\epsilon_0$, and following standard quantization rules, we can directly connect the Lagrangian with the Hamiltonian of Eq.~\eqref{eq_H_low}.

Contrary to a vast majority of work being done in the solid state context, where the Berry curvature is usually defined in the single-particle eigenspectrum (of, e.g., electrons moving within the solid), the above is a \textit{many-body} quantum geometric effect -- by virtue of the coupling to the many-body field $\phi_z$. As such, the energy bands of $H_0$ have no meaningful notion of an occupation number in the second quantized language: the state $\vert 0\rangle$ represents directly a many-body ground state.

\section{\bluehead{Gyration conductance quantization}}\label{sec_gyration_conductance_quantization}

\begin{figure}
\centering
\includegraphics[width=\textwidth]{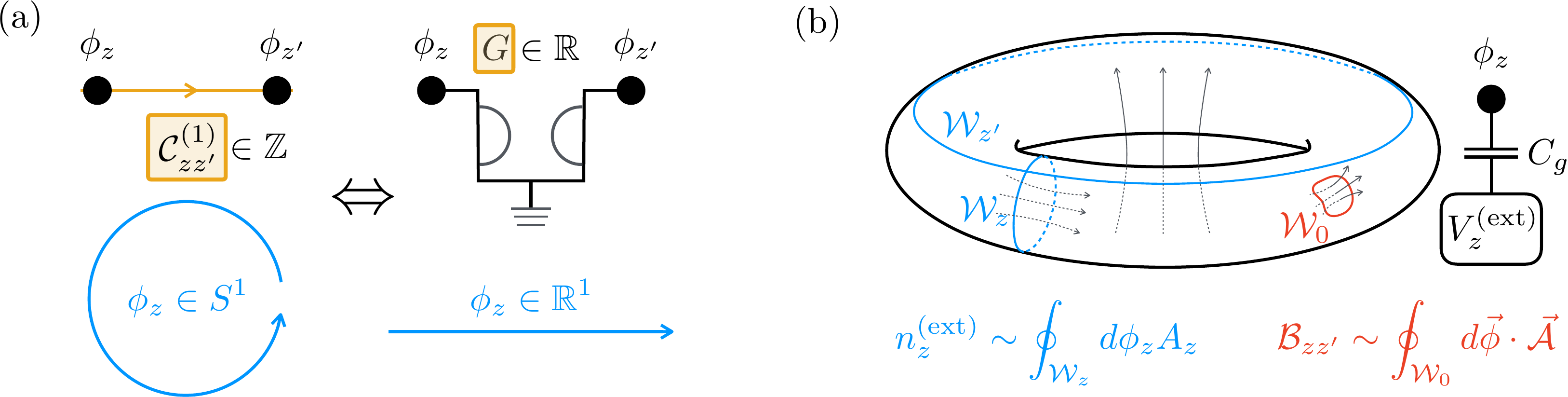}
\caption{Important aspects regarding phase compactness. (a) A nonzero Chern number between two nodes (left) essentially maps onto a gyrator (right, denoted by Tellegen's symbol), however, with the difference that the Chern number (equivalent to the gyration conductance) must be integer quantized when defined on compact superconducting phases. Conceptually, the circuit element is a three-terminal device (one terminal to ground), where in the left symbol, we omit the explicit representation of the ground for simplicity.
(b) Upon including capacitive coupling, the dynamics of a circuit with $Z$ nodes reduces to a fictitious particle moving in a magnetic field on a $Z$-dimensional torus (here shown for two dimensions). Contrary to the ordinary Landau level problem in extended space, the gauge invariance on the torus comes with a number of complications. For instance, in addition to the trivial $\mathcal{W}_{0}$ loops there are two types of nontrivial loops $\mathcal{W}_{z}$ and $\mathcal{W}_{z'}$, which cannot be controlled by the value of the magnetic field only. The phase accumulated across these extra loops is defined by gate-voltage induced offset charges.}
\label{fig:torus}
\end{figure}

Specifically in the quantum circuit context, it was recently noted~\cite{riwar_2023discrete,Virtanen_2024} that geometric effects in the multiterminal Josephson effect are deeply related to quantum circuit versions of gyrator elements~\cite{Viola_2014,Rymarz_2021}, see Fig.~\ref{fig:torus}. The ideal gyrator in Ref.~\cite{Rymarz_2021} is described by a vector potential linear in $\phi_z$. Generalized to an arbitrary number of nodes, it reads
\begin{equation}\label{eq_A_g}
    A_z=\sum_{z^\prime}g_{zz^\prime}\phi_{z^\prime}\ ,
\end{equation}
where the matrix $g$, called gyration conductance matrix, can in principle take arbitrary real values. Note that $g$ is not gauge invariant.
Crucially, though, depending on whether or not we impose compactness of the superconducting phases $\phi_z$, there are further constraints on $g$, rooted in gauge properties on compact spaces. These form the basis of our endeavour.

To begin, note that if we assume $\phi_z$ compact (more precisely, closed---default assumption throughout this work), the first Chern number, defined as
\begin{equation}
    \mathcal{C}^{(1)}_{zz^\prime}=\frac{1}{2\pi}\int_0^{2\pi}d\phi_z\int_0^{2\pi}d\phi_{z^\prime} \mathcal{B}_{zz^\prime}\ ,
\end{equation}
is integer quantized. The quantization itself is the standard first-Chern argument~\cite{Thouless1982,Thouless1983}, which for multiterminal junctions yields a transconductance quantized in units of $4e^2/h$~\cite{Riwar2016}. What is new here is its role as the gyration conductance of a circuit element acting on dynamical, compact phases, Eq.~\eqref{eq_g_matrix}. Compactness also constrains gyrators in general: for any uniform gyration coupling between compact phases, the wave function must be single-valued on the torus, which forces $g-g^T$ to be an integer multiple of $1/2\pi$, just as Dirac's argument quantizes the charge of a magnetic monopole~\cite{simms1976lectures}. A non-integer gyration conductance therefore requires extended phases, whatever the microscopic origin of the gyrator. There is no contradiction with Chern-number cancellation. An isolated
band may have a nonzero Chern number; only the sum over the complete
set of bands must vanish, $\sum_a\mathcal C_a^{(1)}=0.$ Thus the ground-band Chern number is compensated by the excited bands, which are integrated out in the Born--Oppenheimer projection. As indicated in the introduction, nontrivial geometric (Berry curvature) effects, including a resulting nonzero Chern number, have been predicted to emerge for \textit{conventional} multiterminal Josephson junctions~\cite{Riwar2016,klees2021ground,teshler2023ground,Fatemi_2021,Peyruchat_2021,Herrig_2022} with broken time-reversal symmetry. Devices as simple as a three-terminal Josephson junction with a central normal metal scattering region, see Fig.~\ref{fig:qcgyrator}c (left panel), can host a nontrivial Chern number when magnetic flux through the scattering region breaks time-reversal symmetry~\cite{Meyer:2017aa}.

With our above emphasis on the platform independence, we note that the engineering of similar Hamiltonians as in Refs.~\cite{Riwar2016,klees2021ground,teshler2023ground,Fatemi_2021,Peyruchat_2021,Herrig_2022} can also be realized in other systems, e.g., in cavities where the superconducting phase operators $e^{\pm i\phi_z}$ in $H_0$ are replaced with boson ladder operators $b_z^{(\dagger)}$ changing the photon number. For concreteness, in Appendix~\ref{app:cavity_supp}, we present an example of a two-cavity Hamiltonian implementing a Chern insulator, defined in the 2D lattice spanned by the number of photons in the respective cavities, see also the right panel in Fig.~\ref{fig:qcgyrator}c.

In general, all Berry curvatures $\mathcal{B}_{zz^\prime}$, and the potential energy $\epsilon_0$ depend on $\phi_z$ in a periodic (and thus nonlinear) fashion. This means that diagonalizing the low-energy Hamiltonian in Eq.~\eqref{eq_H_low} is still a hard problem. However, we can explicitly separate the topological transition in the form of a non-zero Chern number from any other $\phi_z$-dependent terms, thus ``peeling off'' the topological component of the theory from the otherwise non-linear, strongly correlated physics. To this end, we define $\mathcal{B}_{zz^\prime}=\mathcal{C}_{zz^\prime}^{(1)}/2\pi+\delta \mathcal{B}_{zz^\prime}$ where the integral of $\delta \mathcal{B}_{zz^\prime}$ over $\phi_z,\phi_{z^\prime}$ is zero per definition. This procedure can be reflected by a similar separation of the vector potential into $A_z=a_z+\delta A_z$, where $a_z$ needs to be chosen, such that it returns the nonzero Chern number part when plugged into Eq.~\eqref{eq_Nz_commutator}. Consequently, $a_z$ must be linear, and can be written by means of the gyration conductance, see Eq.~\eqref{eq_A_g}, i.e., $a_z=\sum_{z^\prime}g_{zz^\prime}\phi_{z^\prime}$ with the constraint on the conductance matrix,
\begin{equation}\label{eq_g_matrix}
    \left(g-g^T\right)_{zz^\prime}=\frac{\mathcal{C}^{(1)}_{zz^\prime}}{2\pi}\ .
\end{equation}
We can write $g=g_{\mathrm A}+g_{\mathrm S}$, with $g_{\mathrm A}^T=-g_{\mathrm A}$ and $g_{\mathrm S}^T=g_{\mathrm S}$, the symmetric part contributes only a total derivative, $-\dot{\boldsymbol\phi}^{T}g_{\mathrm S}\boldsymbol\phi=-\frac{d}{dt}\left(\frac12\boldsymbol\phi^Tg_{\mathrm S}\boldsymbol\phi\right)$. It therefore can be gauged away. In physical units, Eq.~\eqref{eq_g_matrix} states that the coupler injects the current $I_z=\sum_{z'}G_{zz'}V_{z'}$ into node $z$, with $G=(2e)^2\mathcal C^{(1)}/h$: the gyration conductance is quantized in units of the Cooper-pair conductance quantum $(2e)^2/h=4e^2/h$, the transconductance quantum of Ref.~\cite{Riwar2016}. Integrated over time, this relates the charge delivered to node $z$ to the node fluxes $\Lambda_{z'}=\int V_{z'}\,dt$, $Q_z=\sum_{z'}G_{zz'}\Lambda_{z'}$. Inverting, a single Cooper pair on node $z'$ corresponds to the flux $\Phi_0\,|q_{zz'}|/p$ at node $z$, with $\Phi_0=h/2e$ and $q_{zz'}/p$ the elements of $(\mathcal C^{(1)})^{-1}$, see Eq.~\eqref{eq_exchange_theta}. This conversion of integer charge into fractional flux underlies the charge--flux response of Sec.~\ref{sec_charge_flux}. The remaining $\delta A_z$ can always be chosen to be periodic in all phases $\phi_z$. The total Hamiltonian can thus be separated into linear and nonlinear parts,
\begin{equation}
    H=H_\text{linear}+H_\text{J}\ ,
\end{equation}
where the linear part reads
\begin{equation}
    H_\text{linear}= 2e^2\sum_{zz^\prime}(n_z+a_z)(C^{-1})_{zz^\prime}(n_{z^\prime}+a_{z^\prime})\ .
\end{equation}
All nonlinear terms, either due to $\delta A_z$ or $\epsilon_0$, are cast into the remaining term $H_J$, which (as we discuss in more detail below) physically represents a generic Josephson effect between nodes. The linear term can be diagonalized exactly. The resulting eigenmodes and emergent excitations of $H_\text{linear}$ remain useful to describe the physics even for finite $H_J$ as long as $H_J$ remains sufficiently small. This is the regime we focus on in what follows: we will show that $H_\text{linear}$ gives rise to bosonic energy eigenstates (generalized Landau levels), where each Landau level has a finitely degenerate subspace. This subspace carries a representation of the projected phase operators. When including a finite $H_J$, the degeneracy lifts, and the projected phase operators enter the low-energy Hamiltonian, giving rise to effects, discussed in Secs.~\ref{sec_charge_flux}--\ref{sec_nonidentical}, some of which map onto known systems and some of which are non-local.

The linear term $a_z$ is not $2\pi$-periodic. A consistent gauge on the torus, linear locally and with jumps that restore periodicity, is constructed in Appendix~\ref{app:gauges_supp}; its Dirac-delta ``horns'' of curvature do not affect the dynamics~\cite{Onofri_2001}.

Overall, $H_\text{linear}$ maps to a particle moving in a constant magnetic field on a $Z$-dimensional torus. Importantly, we stress that the word ``particle'' has to be interpreted with care: as already emphasized, the theory was already formulated in a second-quantized language. The ``positions'' ($\phi_z$) and ``momenta'' ($n_z$) of this particle are already by definition many-body operators. For instance, there exists no physically meaningful way to generalize above torus-analogy to the case of having many particles on a torus. That being said, the image of a particle is nonetheless helpful, as it becomes obvious that the well-known Dirac quantization condition for magnetic monopoles~\cite{simms1976lectures} is perfectly equivalent to the integer quantization of the Chern numbers. We further note that since each Chern number is defined in a 2D subspace, the magnetic field can always be decomposed into components connecting two nodes $z$ and $z^\prime$. It is in this sense, that we can always think of the generic $(Z+1)$-terminal junction ($Z$ nodes plus ground) as a network of simpler fundamental elements connecting two out of the $Z$ nodes, as announced above.

In the extended-phase circuit of Ref.~\cite{Rymarz_2021}, the Josephson potential gives a Hofstadter spectrum in the projected Landau level, which the authors plot against inverse effective flux. In our circuit, compact phases quantize the gyration coupling according to Eq.~\eqref{eq_g_matrix}, and the resulting magnetic translations act within a finite-dimensional Landau level. Other routes to circuit nonreciprocity include the finite-frequency Hall response~\cite{Viola_2014}, reservoir engineering~\cite{Metelmann_2015}, and voltage-tunable junctions~\cite{leroux2022}. Periodic driving has also been proposed to stabilize circuit grid states~\cite{Sellem_2023}.

\section{\bluehead{Landau levels on higher dimensional torus}}\label{sec_LL_torus}

We here diagonalize the Hamiltonian $H_\text{linear}$. It will be convenient to define the kinetic charge $N_z=n_z+a_z$ (which differs from the previously introduced kinetic charge by including only the topological part of the vector potential $a_z$, and not the full $A_z$). If $\phi_z$ were extended, $H_\text{linear}$ would be readily solved through a standard Bogoliubov transformation and yield (infinitely degenerate) Landau levels in a $Z$-dimensional system. Let us briefly recapitulate the diagonalization for extended $\phi_z$, before we delve into the compact case. In this case of the extended $\phi_z$, the kinetic charges $N_z$ can be expressed as linear combinations terms of bosonic ladder operators $\hat a_j^{(\dagger)}$ switching the Landau level number, resulting in the diagonal form, $H=\sum_j \omega_j \hat a_j^\dagger \hat a_j$ (the index $j$ indicates that in more than 2 dimensions, Landau levels have in general many cyclotron eigenfrequencies). At the same time, the Landau-level degeneracy can be captured by zero modes (a set of operators commuting with $N_z$, and thus with $H$), which can be regarded as a kinetic flux,
\begin{equation}
    \Phi_z=\phi_z-\sum_{z^\prime}\left(\frac{1}{g-g^{T}}\right)_{zz^{\prime}}N_{z^{\prime}}\ .
\label{eq:capPhidef}
\end{equation}
This construction of course necessitates that $\det(g-g^T)\neq 0$, in order for $g-g^T$ to be invertible. Furthermore, since $g-g^T$ is skew-symmetric (essentially by definition), the inverse can only exist if $Z$ is even. Both even $Z$ and $\det(g-g^T)\neq 0$ will be the default assumption in the remainder of this work. We will in a moment relate invertibility to the existence of higher Chern numbers.
The non-invertible case is discussed briefly in Sec.~\ref{sec_conclusions_outlook}.

Invertibility requires the coupler to exchange Cooper pairs with ground and therefore has a direct physical meaning. Suppose the coupler has no terminal at ground. A simultaneous shift of all node phases is then implemented on its internal degrees of freedom by a phase-independent charge generator $Q_{\mathrm c}$ according to
\begin{equation}\label{eq:coupler_global_shift}
H_0(\{\phi_z+\lambda\}) = e^{i\lambda Q_{\mathrm c}}H_0(\{\phi_z\})e^{-i\lambda Q_{\mathrm c}}\,.
\end{equation}
The canonical charges $n_z$ generate translations of the phase coordinates, whereas $Q_{\mathrm c}$ acts on the coupler's internal Hilbert space at fixed phases. These are complementary parts of the same symmetry. With the convention in Eq.~\eqref{eq:coupler_global_shift}, the conserved generator of the full Hamiltonian is $\sum_z n_z-Q_{\mathrm c}$, rather than the node charge alone. 
In the locally covariant gauge $|0(\{\phi_z+\lambda\})\rangle =e^{i\lambda Q_{\mathrm c}}|0(\{\phi_z\})\rangle$, each $A_z$ depends only on phase differences and $\sum_z A_z=\langle0|Q_{\mathrm c}|0\rangle \equiv\bar Q_{\mathrm c}$. Eq.~\eqref{eq_Nz_commutator} then yields the gauge-invariant identity
\begin{equation}\label{eq:ground_lemma}
    \sum_z\mathcal B_{zz'}=\partial_{\phi_{z'}}\bar Q_{\mathrm c}\ .
\end{equation}
The right-hand side is the derivative of a periodic function and averages to zero on the torus, while $\mathcal B_{zz'}$ averages to $\mathcal C^{(1)}_{zz'}/2\pi$. Hence $\sum_z\mathcal C^{(1)}_{zz'}=0$ for all $z'$: the vector $(1,\ldots,1)$ lies in the kernel of $\mathcal C^{(1)}$, and $\det\mathcal C^{(1)}=0$. A nonsingular Chern matrix therefore requires a coupler that transfers charge to ground. For $Z=2$ this states that a two-terminal coupler carries no Chern number; the grounded third terminal of the junction in Fig.~\ref{fig:qcgyrator}(c) is essential. Two consequences are used below. First, $\epsilon_0$ and $\delta A_z$ generically contain Fourier harmonics with $\sum_zm_z\neq0$, see Eq.~\eqref{equation_epsilon_Fourier}. Second, the total kinetic charge $N_{\mathrm{tot}}=\sum_zN_z$ is not conserved by the linear part of the Hamiltonian: for $C_{zz'}=C_g\delta_{zz'}$,
\begin{equation}\label{eq:Ntot_not_conserved}
    [H_{\mathrm{linear}},N_{\mathrm{tot}}]=\frac{2ie^2}{\pi C_g}\sum_{z}\Big(\sum_{z'}\mathcal C^{(1)}_{zz'}\Big)N_z\ ,
\end{equation}
which is nonzero whenever $\det\mathcal C^{(1)}\neq0$. The converse does not hold: Eq.~\eqref{eq:Ntot_not_conserved} vanishes only if every row sum of $\mathcal C^{(1)}$ vanishes, and many singular Chern matrices, such as $\mathcal C^{(1)}_{12}=\mathcal C^{(1)}_{13}=1$ with all other elements zero, have nonzero row sums.

Phase compactness has an important impact on the Fock space of our system: the Landau levels have a finite degeneracy. This feature has explicitly been shown for the two-dimensional torus~\cite{Haldane_1985,Onofri_2001} -- a result we here generalize to arbitrary dimensions. To see this, we note that while the construction of the bosonic $N_z$ is still valid on a torus~\footnote{Energy levels on a periodic but extended system and a compact system are related by a Bloch theorem. Therefore, the Landau levels must persist in exactly the same form for both choices.}, we have to construct a set of kinetic phase operators that satisfy $2\pi$-periodicity, which is accomplished with exponentiation, $e^{i\Phi_z}$. These operators still satisfy $[e^{i\Phi_z},N_z]=0$, and provide the algebra
\begin{equation}\label{eq_anyons}
    e^{im\Phi_z}e^{im^\prime\Phi_{z^\prime}}=e^{-i2\pi mm^\prime q_{zz^\prime}/p}e^{im^\prime\Phi_{z^\prime}}e^{im\Phi_z}\ ,
\end{equation}
where $m^{(\prime)}\in\mathbb{Z}$. The sign follows from $[\Phi_z,\Phi_{z'}]=2\pi i\,q_{zz'}/p$, which Eqs.~\eqref{eq:capPhidef} and~\eqref{eq_exchange_theta} give for $[\phi_z,n_{z'}]=i\delta_{zz'}$. We define $\eta_z=e^{i\Phi_z}$ as the elementary projected operators describing the low-energy sector. Because of the second term in Eq.~\eqref{eq:capPhidef}, $\Phi_z$ involves the kinetic charges of all nodes coupled to $z$ through $(\mathcal C^{(1)})^{-1}$. The projected operator $\eta_z$ is therefore collective: its label $z$ identifies the local microscopic term it is projected from, $P_0e^{i\phi_z}P_0\propto\eta_z$, see Eq.~\eqref{eq_Josephson_projected}, not a subsystem on which it acts. This non-locality is a low-energy property: the microscopic couplings are local, and the collective character arises from the projection, much as the ground state of a Hamiltonian with only local couplings is generally non-local. Since $q_{zz'}/p$ is rational, $\eta_z^p$ commutes with all generators and is fixed to a phase in each irreducible sector. Consequently, the $\eta_z$ generate a finite-dimensional projective algebra.

Periodicity of $e^{i\Phi_z}$ is a gauge-fixed statement, whereas the algebra in Eq.~\eqref{eq_anyons} is gauge invariant. For $Z=2$, with $\mathcal C\equiv\mathcal C^{(1)}_{12}$, the symmetric gauge $g=-g^T$ in Eq.~\eqref{eq_A_g} gives $\Phi_1=\phi_1/2+(2\pi/\mathcal C)\,n_2$, so $e^{i\Phi_1}$ is not $2\pi$-periodic in $\phi_1$. In the Landau gauge $a_1=0$, $a_2=-(\mathcal C/2\pi)\,\phi_1$, of the type used in Appendix~\ref{app:gauges_supp} where jumps restore periodicity on the torus, one finds instead
\begin{equation}\label{eq:landau_gauge_translations}
    e^{i\Phi_1}=e^{i2\pi n_2/\mathcal C}\ ,\qquad e^{i\Phi_2}=e^{i\phi_2}\,e^{-i2\pi n_1/\mathcal C}\ .
\end{equation}
Both operators are manifestly $2\pi$-periodic and well defined on integer charges; they are the magnetic translations of Ref.~\cite{Haldane_1985}.

We therefore find that for compact $\phi_z$, the projected operators within a given Landau level obey an abelian exchange algebra, with the exchange phase matrix
\begin{equation}\label{eq_exchange_theta}
    \frac{q_{zz'}}{p} = \frac{1}{2\pi}\left(\frac{1}{g-g^T}\right)_{zz'}\ .
\end{equation}
Since $\mathcal C^{(1)}$ is integer valued, the exchange phase is always a rational number with $q$ being an integer-valued matrix, and $p$ an integer scalar. Explicitly, $p=\mathrm{Pf}\,\mathcal C^{(1)}$, which is the highest Chern number $\mathcal C^{(Z/2)}$ of the device, and $q$ is the matrix of signed Pfaffian minors of $\mathcal C^{(1)}$, the skew-symmetric analogue of the adjugate matrix. Both are therefore integers. The general formulas, a recursion for the Pfaffian and the four-node cyclotron frequencies are given in Appendix~\ref{app:pfaffian}.

To provide specific examples, for two nodes ($Z=2$) we simply have (in matrix representation)
\begin{equation}\label{eq_q/p_2}
    \frac{q}{p}=\frac{1}{\mathcal{C}^{(1)}}\left(\begin{array}{cc}
0 & -1\\
1 & 0
\end{array}\right)\ ,
\end{equation}
where there is only one first Chern number, hence the omission of node indices. In contrast, for four nodes ($Z=4$), we find
\begin{equation}\label{eq_q/p_4}
    \frac{q}{p}=\frac{1}{\mathcal{C}^{(2)}}\left(\begin{array}{cccc}
0 & -\mathcal{C}_{34}^{\left(1\right)} & \mathcal{C}_{24}^{\left(1\right)} & -\mathcal{C}_{23}^{\left(1\right)}\\
\mathcal{C}_{34}^{\left(1\right)} & 0 & -\mathcal{C}_{14}^{\left(1\right)} & \mathcal{C}_{13}^{\left(1\right)}\\
-\mathcal{C}_{24}^{\left(1\right)} & \mathcal{C}_{14}^{\left(1\right)} & 0 & -\mathcal{C}_{12}^{\left(1\right)}\\
\mathcal{C}_{23}^{\left(1\right)} & -\mathcal{C}_{13}^{\left(1\right)} & \mathcal{C}_{12}^{\left(1\right)} & 0
\end{array}\right)\ ,
\end{equation}
where the second Chern number emerges in the denominator, $\mathcal C^{(2)}=\mathcal{C}_{12}^{\left(1\right)}\mathcal{C}_{34}^{\left(1\right)}-\mathcal{C}_{13}^{\left(1\right)}\mathcal{C}_{24}^{\left(1\right)}+\mathcal{C}_{14}^{\left(1\right)}\mathcal{C}_{23}^{\left(1\right)}$. This calculation can be continued in analogy for higher $Z$.

We draw from the above, and from Appendix~\ref{app:pfaffian}, the following conclusions. The requirement $q,p\in\mathbb{Z}$ follows simply from the first Chern numbers $\mathcal{C}_{zz^\prime}^{\left(1\right)}$ being all integer. The existence of the inverse of $g-g^T$ requires $\mathcal{C}^{\left(Z/2\right)}$ (the highest possible Chern number for a device with $Z$ nodes) to be nonzero. Note though, that this by no means implies that $\det(g-g^T)=0$ gives rise to any problematic behaviour whatsoever; in that case one simply has to carefully separate zero and nonzero eigenvalues of $g-g^T$ and construct the kinetic flux operators accordingly, see Sec.~\ref{sec_conclusions_outlook}.

Overall, the above result illustrates that in spite of the system being in principle fully described by the first Chern numbers connecting two nodes, the resulting (nonzero) second and higher Chern numbers appear very prominently in the exchange algebra of multi-node devices. As a matter of fact, all Chern numbers also appear in the energy spectrum. We give the Landau-level frequencies for the case of dominant capacitive coupling ground (for simplicity we assume the capacitance matrix to be $C_{zz^\prime}=C_g\delta_{zz^\prime}$, such that the model only includes a capacitance to ground). For $Z=2$, we obtain a single cyclotron mode, $H=\omega \hat a^\dagger \hat a$, with frequency
\begin{equation}\label{eq_omega_Z2}
    \omega=\frac{2e^2}{\pi C_g}\left|\mathcal C^{(1)}\right|,
\end{equation}
Consequently, the Born--Oppenheimer condition becomes
\begin{equation}
    \frac{\omega}{\Delta_{\min}}
    =
    \frac{
        4E_C\left|\mathcal C^{(1)}\right|
    }{
        \pi\Delta_{\min}
    }
    \ll1,
    \qquad
    E_C=\frac{e^2}{2C_g}. \label{eq:BO_validity_Z2}
\end{equation}
For orientation, aluminium has $\Delta_{\mathrm{Al}}/h\approx44$\,GHz. Since $\mathcal C^{(1)}$ can change only where the coupler gap closes, $\Delta_{\min}$ is set by the distance of the operating point from the nearest such transition, for instance a Weyl point in the parameter space of Ref.~\cite{Riwar2016}. Assuming this distance to be a sizeable fraction of the superconducting gap, $\Delta_{\min}/h\approx5$--$15$\,GHz, and $E_C/h\approx0.3$--$1$\,GHz as in transmon-type circuits, Eq.~\eqref{eq:BO_validity_Z2} with $|\mathcal C^{(1)}|=1$ gives $\omega/h\approx0.4$--$1.3$\,GHz and $\omega/\Delta_{\min}\approx0.03$--$0.25$. We therefore expect the Born-Oppenheimer condition to be accessible. In the Cooper-pair-box regime $E_C$, and with it $\omega$, can be considerably larger, which tightens the condition. The projection also renormalizes the nonlinear terms of $H_J$, see Eq.~\eqref{eq_Josephson_projected} below. For a few nodes this renormalization is weak; we expect it to grow with the number of nodes, which will be addressed in future work.
For general $Z$, the frequencies are $\omega_j=(2e^2/\pi C_g)\,\lambda_j$, where $\pm i\lambda_j$ are the eigenvalues of $\mathcal C^{(1)}$. For $Z=4$ they depend on both the first and the second Chern numbers, see Appendix~\ref{app:pfaffian}.
As a consequence, both second and first Chern numbers can be directly tracked in an experiment measuring the circuits resonance frequencies (such as, e.g., a standard dispersive readout).
Let us return to the degenerate subspace within a Landau level. Given that the exchange phase is rational, the algebra in Eq.~\eqref{eq_anyons} admits finite-dimensional projective
representations, since $e^{ip\Phi_z}$ commutes with all generators, and thus must return an ordinary c-number (consistent with one of the fundamental properties of parafermion operators~\cite{Fendley_2012}).
 The general case is settled by the normal form given below, Eq.~\eqref{eq:frobenius}.
At any rate, for the special cases we consider in the remainder of this work it is easy to find explicit representations, where each Landau level is exactly $p$-fold degenerate. The simplest case that can be computed exactly is that of two nodes, mapping onto the 2D torus, where $p=\mathcal{C}^{(1)}$. In the regular Landau problem (on an extended 2D space), the level degeneracy is given by the ratio of the sample area divided by the size of the corresponding cyclotron orbit. On the torus, essentially the same logic remains true, leading to a $p=\mathcal{C}^{(1)}$ fold degeneracy, see also Fig.~\ref{fig:gauge} in Appendix~\ref{app:gauges_supp}.

The degeneracy and the structure of the algebra follow in general from a normal form. For every integer skew-symmetric $\mathcal C\equiv\mathcal C^{(1)}$ with $\det\mathcal C\neq0$ there is a unimodular integer matrix $U$, that is, a $Z\times Z$ matrix with integer entries and $\det U=\pm1$, such that
\begin{equation}\label{eq:frobenius}
    U^{T}\mathcal C\,U=\bigoplus_{i=1}^{Z/2}d_i\begin{pmatrix}0&1\\-1&0\end{pmatrix},\qquad d_1\,|\,d_2\,|\,\cdots\,|\,d_{Z/2}\ ,\qquad \prod_id_i=|\mathrm{Pf}\,\mathcal C|\ .
\end{equation}
This is the normal form of skew-symmetric integer matrices~\cite{Newman1972}. Writing $W(\mathbf m)=e^{i\mathbf m\cdot\boldsymbol\Phi}$ for $\mathbf m\in\mathbb Z^Z$, Eqs.~\eqref{eq_anyons} and~\eqref{eq_exchange_theta} read $W(\mathbf m)W(\mathbf n)=e^{-2\pi i\,\mathbf m^T\mathcal C^{-1}\mathbf n}\,W(\mathbf n)W(\mathbf m)$, so two projected processes commute if and only if $\mathbf m^T\mathcal C^{-1}\mathbf n\in\mathbb Z$. The operators $W'_k=W(U^{-T}\mathbf e_k)$ obey the same relations with $\mathcal C$ replaced by its block-diagonal form. The algebra is therefore a product of $Z/2$ independent clock--shift pairs of dimensions $d_i$, whose $d_i$-th powers are central and fixed to phases by the offset charges, and its irreducible representations have dimension $|\mathrm{Pf}\,\mathcal C|$. This equals the number of states in each Landau level, as given by the Riemann--Roch index theorem on $\mathbb T^Z$~\cite{Mumford1970,BirkenhakeLange2004} (see Refs.~\cite{Onofri_2001,Haldane_1985} for the two-dimensional torus). Each Landau level thus carries exactly one irreducible representation, and the $|\mathrm{Pf}\,\mathcal C|$-fold degeneracy holds for all $Z$, not only for the cases treated explicitly. The columns of $U^{-T}$ specify which products of node operators $\eta_z$, that is, which multi-node Josephson harmonics, realise each clock--shift pair. A sparse Chern matrix generally yields a dense $\mathcal C^{-1}$. For instance, the chain $\mathcal C_{12}=\mathcal C_{34}=2$, $\mathcal C_{23}=1$ has no coupler between nodes $1$ and $4$, yet $\mathrm{Pf}\,\mathcal C=4$, $(d_1,d_2)=(1,4)$, and Eq.~\eqref{eq_q/p_4} gives $q_{14}/p=-1/4$.

A finite $H_J$ lifts this degeneracy, see Sec.~\ref{sec_charge_flux}, but leaves the algebra of Eq.~\eqref{eq_anyons} intact as long as it does not mix Landau levels. The Chern numbers that fix the algebra are stable under any deformation of the coupler that keeps its gap open; they change only when a Weyl point of the coupler spectrum is crossed, so that moving the Weyl points, for instance with a loop flux, tunes the Chern matrix and hence the exchange algebra in situ~\cite{Riwar2016}. Processes that change the ground state of the coupler, chiefly quasiparticle poisoning, change the Chern numbers and are assumed to be slow on the time scale of the experiment.

\section{\bluehead{Uncommon gauge properties: integer charges and fractional fluxes}}\label{sec_charge_flux}

 The exchange algebra of Eq.~\eqref{eq_anyons} has a measurable counterpart in the response to external fields. We will show in the following, that when sticking to the quantum circuit interpretation of the field $\phi_z$, the flux periodicity is reduced to $\Phi_0/p$ while charge stays integer, in contrast to the fractional quantum Hall effect, and that both flux fractionalization and exchange algebra can be probed by a finite $H_J$. The way external fields enter is fixed by the microscopic circuit; its relation to gauge invariance is discussed in Appendix~\ref{app:gauge_fields}.

Concerning the nonzero $H_J$, note that both $\epsilon_0$ and $\delta A_z$ are $2\pi$-periodic in $\phi_z$, and can therefore be expressed in terms of a generic Fourier series of harmonics, i.e.,
\begin{equation}\label{equation_epsilon_Fourier}
\epsilon_0=\sum_{m_1,m_2,\ldots,m_Z} \epsilon_{m_1,m_2,\ldots,m_Z}e^{i m_1\phi_1}e^{i m_2\phi_2}\ldots e^{i m_Z\phi_Z}
\end{equation}
with $m_z\in\mathbb{Z}$ (and similarly for $\delta A_z$). If this Josephson effect is weak compared to the cyclotron frequency, we can project onto the lowest Landau level with the operator $P_0$ (such that $\hat a_jP_0=P_0\hat a_j=0$ for all cyclotron modes $j$, while the degenerate subspace stays untouched). For the first harmonic, we get
\begin{equation}\label{eq_Josephson_projected}
    P_0e^{i\phi_z}P_0= \alpha e^{i\Phi_z}\ ,
\end{equation}
where the scalar prefactor is real, smaller than $1$, and equal to $\alpha=\text{tr}[e^{i\sum_{z^\prime}\left(\frac{1}{g-g^{T}}\right)_{zz^{\prime}}N_{z^{\prime}}}P_0]$. It simply arises from projecting out the Landau level excitations, which renormalizes the amplitude of the Josephson effect. The factor $\alpha$ depends on the Chern matrix and on the shape of the capacitance matrix, but not on its overall scale, so it survives the limit of vanishing capacitance. A reduction that drops the capacitive energy before quantizing, keeping $L_G$ alone as a first-order Lagrangian, gives $\alpha=1$ and misses this factor: here, reduction and quantization do not commute. Dunne, Jackiw and Trugenberger found the same effect for a harmonic potential, where it appears as a finite shift of the zero-point energy~\cite{Dunne_1990}. For the Josephson potential, it renormalizes $E_J$.
The same principle generalizes for arbitrary higher harmonics,
\begin{equation}\label{eq_Josephson_projected_general}
    P_0 e^{i\sum_z m_z \phi_z}P_0\sim e^{i\sum_z m_z \Phi_z}\ .
\end{equation}
Note that we also get the same qualitative result for terms due to $\delta A_z$, which in lowest order provide extra terms of the form $P_0 N_{z\prime} e^{i\sum_z m_z \phi_z}P_0\sim e^{i\sum_z m_z \Phi_z}$. The only difference here is that the renormalization prefactor changes shape due to the extra $N_z$ operator.

\begin{figure}
\includegraphics[width=\textwidth]{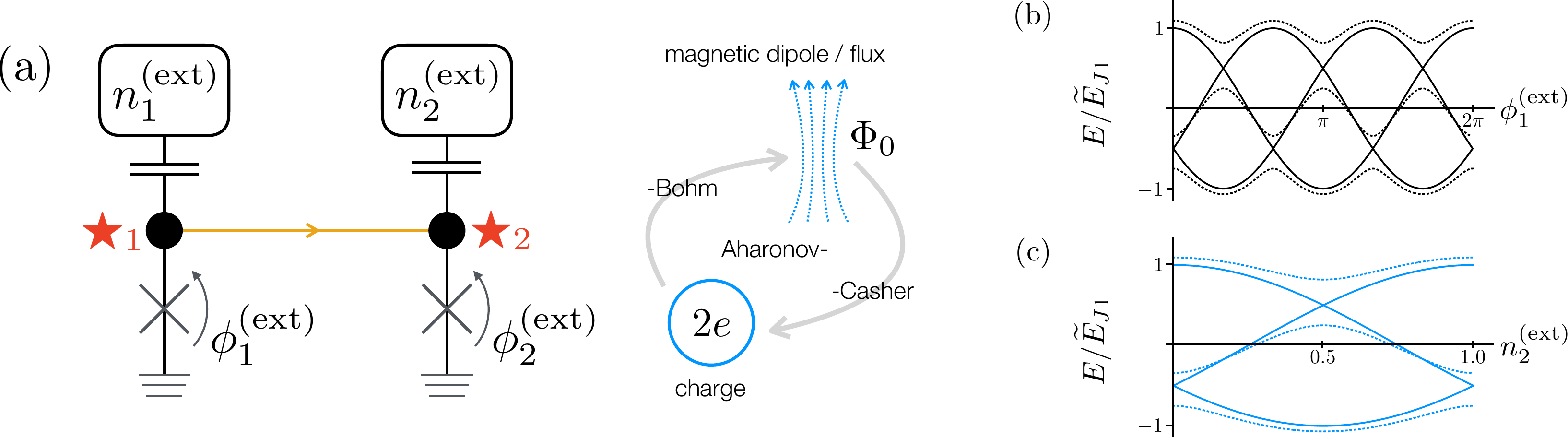}
\caption{Gapping the degeneracy of the lowest Landau level by means of Josephson couplings reveals the intricate gauge properties on the torus, and the resulting nonlocal interplay between charge and flux. (a) Minimal example of a circuit with two nodes, connected by a Chern number, each additionally coupled to ground via a Josephson junction. Coupling to an external magnetic field is represented by flux shifts for each Josephson junction, $\phi_{1,2}^{\left(\text{ext}\right)}$, and interaction to a static electric field is accounted for by a capacitive coupling to a gate voltage, resulting in the offset charges $n_{1,2}^{\left(\text{ext}\right)}$. The external fields result in a finite flux and charge dispersion, which can be interpreted as a superposition of an Aharonov-Bohm, respectively, Aharonov-Casher effect. Panels (b) and (c) show the different periodicity with respect to applied flux at node 1 and applied offset charge at node 2, here for $p=\mathcal C^{(1)}=3$. The broken periodicity with respect to the offset charge indicates that the projected operators shift the flux by fractions of $\Phi_0$ but the charge by integers. The solid lines show the case when only one site is shunted by a Josephson junction (i.e.\ $\widetilde{E}_{J2}=0$), while the dashed lines mark the case when $\widetilde{E}_{J2}=\frac12 \widetilde{E}_{J1}$. As sets, the spectra are $2\pi/p$-periodic in $\phi_1^{(\mathrm{ext})}$ and $1$-periodic in $n_2^{(\mathrm{ext})}$; the $p$-fold charge period refers to the individual solid branches (see text).
}
\label{fig:spectrum}
\end{figure}

Overall, a finite $H_J$ lifts the degeneracy of the Landau level. For an explicit example, let us add a regular Josephson effect between ground and node $z$, $-E_J\cos(\phi_z)$. Projected to the lowest Landau level, this yields the Hamiltonian $H=-\widetilde{E}_J\cos(\Phi_z)$, revealing a fine structure of the degenerate subspace. The Josephson energy is renormalized, $E_J\rightarrow \alpha E_J\equiv \widetilde{E}_J$ in accordance with Eq.~\eqref{eq_Josephson_projected}. For example, for $Z=2$ the prefactor is simply $\alpha=e^{-\pi/(2p)}$ (i.e., the higher the value of the first Chern number $\mathcal{C}^{(1)}$, the weaker the renormalization). A Josephson effect between two nodes $\sim\cos(\phi_z-\phi_{z^\prime})$ is treated on the exact same footing, yielding at low energies the projected tunneling term, $\sim\cos(\Phi_z-\Phi_{z^\prime})$. This is the projected Cooper-pair tunneling between the two nodes. 

Importantly, the resulting energy splitting is affected by the coupling of external electromagnetic fields, which, in the circuit context are known as Aharonov-Bohm \textit{and} Aharonov-Casher effects~\cite{Elion1993,Cottet_2002,Mooij_2006,Pop2012,Astafiev_2012,Koliofoti_2023}. The inclusion of both applied voltages (Aharonov-Casher) and magnetic fluxes (Aharonov-Bohm) requires some attention to detail. The latter is in principle straightforwardly included as a classical flux offset, $\phi_z\rightarrow \phi_z+\phi_z^\text{(ext)}$, which arises from a coupling to an external magnetic field~\cite{Riwar_2022}. However, the question is, whether this shift in $\phi_z$ has to be applied to all parts in the Lagrangian, whenever $\phi_z$ appears, only to selected terms (e.g., we could shift all $\phi_z$ by an offset within $\epsilon_0$, while gyration terms $\sim \phi_z\dot{\phi}_{z^\prime}$ are unchanged). The answer is that it depends on the physical realization of the circuit. In the following, we will apply the shift only to Josephson effects, and justify this choice after the fact.

The Aharonov-Casher effect emerges when including an offset charge to node $z$, $n_z\rightarrow n_z+n_z^{(\text{ext})}$, by applying the gate voltage $V_{gz}$ to node $z$, such that $n_z^{(\text{ext})}=C_{gz}V_{gz}/(2e)$, where $C_{gz}$ is the capacitance between node $z$ and the gate attached to it.
It is well known that for circuits with compact phases (such as the charge qubit~\cite{Cottet_2002}), the energy spectrum is sensitive to charge offsets---giving rise to the so-called charge dispersion effect (which was, e.g., recently used to track quasiparticle poisoning~\cite{Serniak2018,Serniak2019} or to probe higher Josephson harmonics~\cite{Willsch_2024}, indicating that offset charge noise is slow enough to allow reliable measurements). For external voltage sources, such offsets can be added readily in the capacitive term of the original microscopic Hamiltonian, Eq.~\eqref{eq_including_capacitances}, which is then simply carried over to the low energy theory, Eq.~\eqref{eq_H_low}. Note however, that the quantum geometric component may have its own contribution. In quantum mechanical terms, quantum systems on compact manifolds pick up a phase as the coordinate travels along the periodic $\phi_z$ and completes one cycle. For a two-dimensional subspace ($\phi_z,\phi_{z^\prime}$), this means that the system is not fully determined by the Berry curvature $\mathcal{B}_{zz^\prime}$ alone, since it only determines the geometric phases of paths that do not probe the torus topology (i.e., paths that can be shrunk to single points). The torus however offers two more distinct types of paths (see Fig.~\ref{fig:torus}) whose phases need to be fixed, thus nominally contributing to $n_z^{(\text{ext})}$.

At this stage, we can already see that there is a peculiar interconnectedness between charge and flux offsets: the Chern number part of the vector potential depends on $\phi_z$, see Eq.~\eqref{eq_A_g}, such that offset charges can be swallowed by offset fluxes and vice versa.
Let us focus on pure external voltage sources [see also Fig.~\ref{fig:spectrum}(a)], and adding phase offsets as discussed previously. Then, the respective flux and charge offsets carry over to the kinetic quantities $N_z$ and $\Phi_z$ respectively. When projecting to the lowest Landau level, the Josephson effect thus reveals both types of shifts as
\begin{equation}\label{eq_Josephson_effect}
    H=-\widetilde{E}_J\cos\bigg(\Phi_z+\underbrace{\phi_z^{(\text{ext})}-2\pi\sum_{z^\prime}\frac{q_{zz^\prime}}{p}n_{z^\prime}^{(\text{ext})}}_{\Phi_z^{(\text{ext})}}\bigg)\ ,
\end{equation}
where we defined the total phase shift $\Phi_z^{(\text{ext})}$ as the sum of flux and charge offsets. Since the eigenvalues of $e^{i\Phi_z}$ are discrete and bounded, neither flux nor charge offsets can be gauged away (unless they take on precise quantized values). Instead, in the low-energy projection, the circuit's eigenspectrum is explicitly sensitive to both the Aharonov-Bohm phase (regular Josephson effect) and the Aharonov-Casher phase (effectively a dual Josephson effect), with the latter being \textit{fractional}, as it directly reflects the exchange algebra. This is in contrast to the regular fractional Josephson effect, where the periodicity in the Aharonov-Bohm phase is altered, (e.g., the $4\pi$-periodic dependence of Majorana-based junctions on the Aharonov-Bohm phase). The two effects are nonetheless closely related. Both enlarge the period of individual branches, protected by a conserved label, as discussed after Eq.~\eqref{eq_H_twonodes_external_fields}.

This demonstrates that the projected operators shift the charge by integers and the flux by fractions of $\Phi_0$, and both shifts enter in superposition. This duality
can be better understood if we perceive the flux as a magnetic dipole~\cite{Elion1993}, where we cannot distinguish whether the charge is moving around the flux, or vice versa. We can detect only a mutual movement of charge and flux with respect to each other [as illustrated in Fig.~\ref{fig:spectrum}(a)]. Overall, the duality between charge and phase manifests itself in Eq.~\eqref{eq_Josephson_effect} as a highly nonlocal interdependency between applying a flux offset between ground and node $z$ and adding a gate (inducing an offset charge) at another node $z^\prime$ defined by the connectivity of the gyrator network.\footnote{This interdependence results from eliminating the propagation of electromagnetic waves, which is valid when the speed of light is fast compared with all other time scales, as is standard in circuit QED~\cite{Riwar_2022}. Within the Born--Oppenheimer limit, the response time of the coupler, of order $\hbar/\Delta_{\min}$, must be short as well. Flux noise in a dc-SQUID can broaden resonance lines through fluctuations in the tunable Josephson energies. The size of this effect depends on the flux sensitivity at the operating point and the flux-noise spectrum over the relevant measurement bandwidth~\cite{Anton2013}.}

Let us illustrate both the fractionalization and interdependency of charge and phase explicitly for $Z=2$. The exchange algebra with the phases of Eq.~\eqref{eq_q/p_2} is here represented, e.g., by the constructions
\begin{equation}\label{eq_qudit_Z2}
    e^{i\Phi_1}=Z_p,\quad e^{i\Phi_2}=X_p.
\end{equation}
where $Z_p,X_p$ form the unitary operator basis for a \textit{qudit} with $p$ degenerate states $\{|d\rangle\}$ ($d=1,2,\ldots,p$), such that $Z_p |d\rangle=e^{i2\pi d/p}|d\rangle$ and $X_p|d\rangle =|d+1\mod p\rangle$. These generators obey $Z_pX_p=e^{i2\pi/p}X_pZ_p$, and therefore realize precisely the finite magnetic-translation, or clock-shift, algebra.
We now add a Josephson effect according to Eq.~\eqref{eq_Josephson_effect} to each of the two nodes (with Josephson energies $\widetilde{E}_{J1}$ and $\widetilde{E}_{J2}$). Note that this Josephson effect could have various origins: it could correspond to extra Josephson junction that is connected to a circuit containing an ideal gyrator (without intrinsic Josephson effect), or it could belong to the gyrating element, in case the latter is not ideal (and thus contains spurious Cooper-pair leakage). Using the representation in Eq.~\eqref{eq_qudit_Z2}, we get the Hamiltonian
\begin{equation}\label{eq_H_twonodes_external_fields}
\begin{split}
    H=-\widetilde{E}_{J1}\cos\left(\Phi_1+\phi_1^{\text{(ext)}}+\frac{2\pi}{p}n_2^{\text{(ext)}}\right)\\-\widetilde{E}_{J2}\cos\left(\Phi_2+\phi_2^{\text{(ext)}}-\frac{2\pi}{p}n_1^{\text{(ext)}}\right)\ ,
\end{split}
\end{equation}
which is easy to diagonalize.
The resulting energy spectrum is shown in Fig.~\ref{fig:spectrum}(b,c). It exhibits flux-fractionalization in the sense of a broken periodicity with respect to offset charge, depending on the value of $p$.

This statement concerns individual branches and should be read precisely. Conjugation by $X_p=e^{i\Phi_2}$ shifts $\Phi_1$ by $2\pi/p$, which in Eq.~\eqref{eq_H_twonodes_external_fields} is equivalent both to $n_2^{(\mathrm{ext})}\to n_2^{(\mathrm{ext})}+1$ and to $\phi_1^{(\mathrm{ext})}\to\phi_1^{(\mathrm{ext})}+2\pi/p$. For any $\widetilde E_{J2}$, the spectrum as a set is therefore $1$-periodic in $n_2^{(\mathrm{ext})}$ and $2\pi/p$-periodic in $\phi_1^{(\mathrm{ext})}$; in particular, so is the ground-state energy. A clock-labelled branch returns to itself only after $p$ units of offset charge, and only for $\widetilde E_{J2}=0$: for $\widetilde E_{J2}\neq0$ the crossings become avoided [dashed lines in Fig.~\ref{fig:spectrum}(c)], and the $p$-fold period requires diabatic passage through the gaps. The reduced flux period, by contrast, is robust and generalises. Conjugation by $W(\mathbf m)$ shifts $\boldsymbol\Phi$ by $\pm2\pi\mathcal C^{-1}\mathbf m$. The choice $\mathbf m=(\chi/2\pi)\,\mathcal C\,\mathbf e_z$, which is integer for $\chi\in(2\pi/g_z)\mathbb Z$ with $g_z=\gcd_{z'}\mathcal C^{(1)}_{z'z}$, shifts $\Phi_z$ alone. Whenever $\Phi_z$ enters the projected Hamiltonian only through a flux-biased junction, the spectrum is thus periodic in $\phi_z^{(\mathrm{ext})}$ with period $2\pi/g_z$: a junction shunting node $z$ of a Chern network has flux period $\Phi_0/g_z$, with $g_z=p$ for two nodes.

The two periodicities of Eq.~\eqref{eq_H_twonodes_external_fields} are two faces of one effect. Per branch, the flux dependence is an ordinary $2\pi$-periodic Josephson relation, whereas the offset-charge dependence has period $p$, that is, $2pe$ in gate charge. This is the fractional dual Josephson effect: the Aharonov--Casher phase of a Cooper pair encircling the flux $\Phi_0/p$ that, by the gyrator relation of Sec.~\ref{sec_gyration_conductance_quantization}, a single Cooper pair on node 2 induces at node 1. As for the $4\pi$-periodic Josephson effect of Majorana junctions~\cite{Kitaev_2001,Badiane:2013aa,Wiedenmann_2016}, the enlarged period belongs to individual branches and is protected by a conserved label. There, the label is fermion parity. Here, for $\widetilde E_{J2}=0$, it is the $\mathbb Z_p$ eigenvalue of $e^{i\Phi_1}$, which then commutes with $H$. Any term containing $\Phi_2$, such as the junction at node 2, breaks this label, turns the crossings into avoided crossings and restores the period of one Cooper pair, just as quasiparticle poisoning restores $2\pi$-periodicity in the Majorana case. For $p=2$ the correspondence is exact: the branches
\begin{equation}\label{eq:dual_majorana}
    E_\pm=\mp\widetilde E_{J1}\cos\!\left(\phi_1^{(\mathrm{ext})}+\pi n_2^{(\mathrm{ext})}\right)
\end{equation}
are the charge dual of the Majorana spectrum $\pm E_M\cos(\phi/2)$, with period $4e$ in gate charge instead of $4\pi$ in phase. The spectrum as a set, by contrast, is $2\pi/p$-periodic in flux, so the current--phase relation of the shunt contains only harmonics of order $p$, as for correlated transfer of $p$ Cooper pairs; for $p=2$ the period is $\Phi_0/2$. Related multiterminal-junction experiments have probed Cooper-quartet transport~\cite{pfeffer_subgap_2014,cohen_nonlocal_2018} and observed half-flux-quantum interference in a quartet flux loop~\cite{Huang_2022}. There is thus no fractional Josephson effect on the flux side: the fractionalization resides entirely in the charge dependence. For general $Z$, an offset charge at node $z'$ shifts the argument of a junction at node $z$ by $2\pi q_{zz'}/p$ per Cooper pair, see Eq.~\eqref{eq_Josephson_effect}. When no other term fails to commute with $e^{i\Phi_z}$, a branch therefore returns after $p/\gcd(p,q_{zz'})$ Cooper pairs. In the chain example of Sec.~\ref{sec_LL_torus}, a gate on node 4 produces a fractional dual Josephson effect with a period of four Cooper pairs at a junction on node 1, although no coupler joins the two nodes.

 The superposition of fractional flux and integer charge is further demonstrated by the fact that offset charges at node $1$ and the offset fluxes at node $2$ are linearly dependent. Which offsets enter the projected Hamiltonian, and how, depends on where the junctions and loops sit in the circuit. With the gyrator as a separate element coupled to ground, the two junctions in Fig.~\ref{fig:spectrum}(a) form two distinct loops, so that $\phi_1^{(\mathrm{ext})}$ and $\phi_2^{(\mathrm{ext})}$ act separately; a Josephson coupling intrinsic to the gyrating element would form no loop, and the flux dependence would disappear. The relation to gauge invariance is discussed in Appendix~\ref{app:gauge_fields}.

The term flux fractionalization also appears for half-quantum and fractional vortices~\cite{rampp2022integer,geshkenbein1987vortices,kee2000half,ivanov2001non} and in kagome superconductor rings~\cite{ge2024charge}. There it denotes a flux quantum below $h/2e$ that reflects the structure or the charge of the condensate. We observe the converse: the condensate is unchanged, and the flux that a single Cooper pair induces through the gyrator, $\Phi_0/p$ for two nodes, is fractional, which becomes visible under an applied electric rather than magnetic field.

\section{\bluehead{Mapping to fermions}}\label{sec_simulation_fermion}

One particularly promising application for quantum hardware is the simulation of fermionic quantum many-body systems, typically described as
\begin{equation}\label{eq_fermion_H}
    H=\sum_{z_1 z_2}h_{z_1z_2}^{(2)}c^\dagger_{z_1}c_{z_2}+\!\!\sum_{z_1 z_2 z_3 z_4}\!\!h_{z_1z_2z_3z_4}^{(4)}c_{z_1}^\dagger c_{z_2}c_{z_3}^\dagger c_{z_4},
\end{equation}
where the fermion operators $\{c_{z},c^{\dagger}_{z^\prime}\}=\delta_{zz^\prime}$ are cast into an appropriate basis. Such Hamiltonians are of great interest for quantum chemistry computations~\cite{Cao_2019,McArdle_2020}, but also various solid state problems~\cite{Barends_2015,Kaicher_2020,Khodaeva_2024}. Generalizations from regular fermions to Majoranas further provide interesting connections to the SYK model~\cite{Sachdev_1993,Kitaev_2015} and host so-called fracton excitations~\cite{you2019building,pretko2020fracton}.

Qubit-based quantum computational chemistry has been widely studied, and is considered one of the prime applications of near-term quantum hardware~\cite{Cao_2019,McArdle_2020}. While a Fock space representation allows for a straightforward state initialization~\cite{Abrams_1999}, there usually remains the well-known problem of fermion statistics. The anticommutation already causes problems for classical computation methods such as the famous sign problem in quantum Monte Carlo methods~\cite{Troyer_2005}. In digital quantum simulation~\cite{Lloyd_1996}, a Jordan--Wigner mapping can represent fermionic terms by long Pauli strings~\cite{Whitfield_2011}. Implementing those strings with elementary gates can increase circuit depth, a resource constrained by noise on near-term hardware~\cite{Preskill_2018}. As a consequence, the nonlocality of Jordan-Wigner chains has led to the development of more efficient (but still nonlocal) encodings of anticommutation properties~\cite{Bravyi_2002,Havlicek_2017} or to sophisticated techniques for increasing the simultaneity of qubit gates~\cite{Christandl_2004,Yung_2006,Vinet_2012,Chapman_2016,Naegele_2022}.

\begin{figure}[t]
\centering
\includegraphics[width=0.5\textwidth]{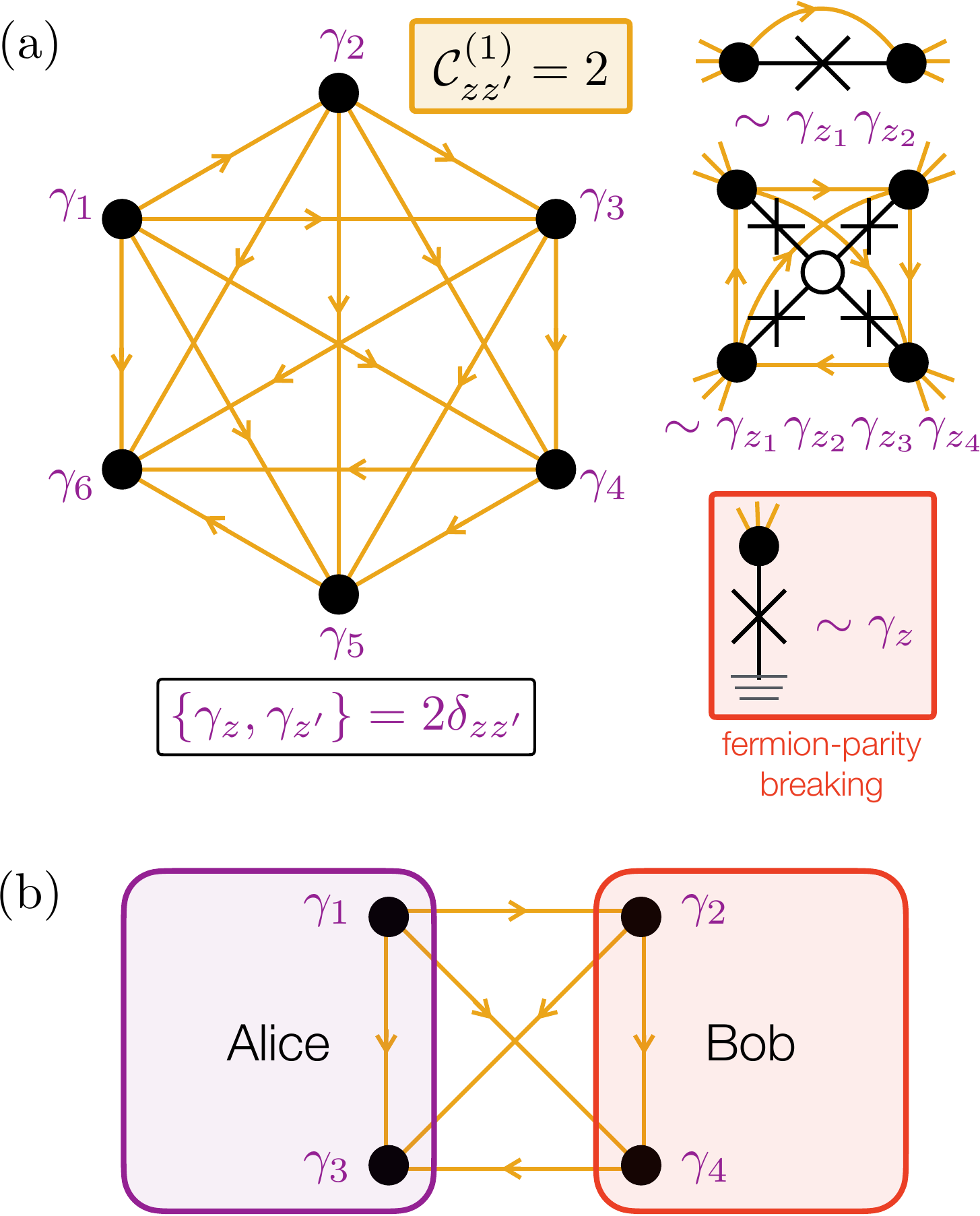}
\caption{Mapping from Chern networks to fermionic systems. (a) Choosing all Chern numbers identically equal 2 (in a complete graph) reduces the projected operators to encoded Majorana operators. Quadratic and quartic interaction terms are equivalent to two-terminal, respectively 4-terminal Josephson couplings. Note that with Josephson couplings to ground (which do not conserve the total Cooper pair charge on the nodes) the circuit architecture allows for the generation of terms in the Hamiltonian which break the fermion-parity superselection rule. (b) Fermion-parity breaking allows protocols that evade the no-communication argument for local fermionic modes. It can be probed by two participants, Alice and Bob, who each act only on their own nodes.}
\label{fig:ifermion}
\end{figure}

In contrast, the here presented many-body quantum geometric mechanism provides a direct mapping from scalar fields to fermionic Hamiltonians [Eq.~\eqref{eq_fermion_H}], which permits the use of \textit{local} interactions and control.

To start, take a generic circuit with $Z$ nodes (recall that $Z$ is even), and set each first Chern number equally to the value $2$, i.e., $\mathcal{C}_{zz^\prime}^{(1)}=2$ (for $z^\prime >z$). In accordance with Eq.~\eqref{eq_exchange_theta} and the subsequent discussion, we here find an exchange algebra with $p=2^{Z/2}$ and (again with $z^\prime >z$) $q_{zz^\prime}=(-1)^{z+z^\prime}2^{Z/2-1}$ [where $(-1)^{z+z^\prime}$ simply provides an alternating sign depending on the $z$ and $z^\prime$ indices, due to the Levi-Civita tensor in Eq.~\eqref{eq_q_matrix_app}]. Ultimately, all pairs of nodes need to be connected to form a complete graph, see, e.g., Fig.~\ref{fig:qcgyrator}. We thus observe the following. The lowest Landau level has $p=2^{Z/2}$-fold degeneracy. Moreover, all exchange ratios simplify to $q_{zz^\prime}/p=\pm 1/2$, such that the projected operators $e^{i\Phi_z}$ anticommute, and for the same reason $(e^{i\Phi_z})^2=e^{i2\Phi_z}$ must be a c-number (which, upon appropriate choice of the offset phase contributions, can be set to 1). The system thus generates a set of $Z$ real Majorana fermions,
\begin{equation}\label{eq_gamma_z}
    e^{i\Phi_z}\equiv \gamma_z
\end{equation}
with $\{\gamma_z,\gamma_{z^\prime}\}=2\delta_{zz^\prime}$. Since $Z$ is even, we can always represent these $Z$ Majoranas in terms of $Z/2$ ordinary complex fermions, and thus open the road towards implementations of the Hamiltonian in Eq.~\eqref{eq_fermion_H}. To this end, we introduce the well-known mapping between the Majorana basis and the actual fermion basis, by choosing the representation
\begin{equation}
\label{eq_gamma_vs_c}
\begin{split}
    \gamma_z&=c_z+c^\dagger_z,\\
    \gamma_{z+Z/2}&=i(c_z-c_z^\dagger)\ ,
\end{split}
\end{equation}
where the $z$ index range is cut in half, $z=1,\ldots,Z/2$, when referring to fermions instead of Majoranas. This mapping is not unique. Any $\mathrm{SO}(Z)$ transformation applied to $\gamma_{z}$ still results in the same fermionic algebra.

In order to get a nonzero coupling between the fermions, i.e., to get to the actual form of Eq.~\eqref{eq_fermion_H}, we again need a nonzero $H_J$. Importantly, in order to get to a regular fermionic field, we need to \textit{restrict} $H_J$. Namely, we restrict $H_J$ to the case where $\delta A_z$ and $\epsilon_0$ (which, as a reminder, are both periodic in $\phi_z$) depend only on phase differences, $\phi_z-\phi_{z^\prime}$. In that case, the only Fourier components in $\epsilon_0$ and $\delta A_z$, see also Eq.~\eqref{equation_epsilon_Fourier}, are the ones where $\sum_z m_z =0$. If we now project $H_J$ onto the lowest Landau level $e^{i\phi_z}\rightarrow e^{i\Phi_z}$ (as developed in the previous section), and use above mapping to Majoranas, we arrive at a Hamiltonian of the form
\begin{equation}\label{eq_H_Majorana}
    H=\sum_{z_1 z_2}\widetilde{h}_{z_1z_2}^{(2)}\gamma_{z_1}\gamma_{z_2}+\!\!\sum_{z_1 z_2 z_3 z_4}\!\!\widetilde{h}_{z_1z_2z_3z_4}^{(4)}\gamma_{z_1}\gamma_{z_2}\gamma_{z_3}\gamma_{z_4}+\ldots \ .
\end{equation}
 Only even terms appear because of a selection rule. A harmonic $e^{i\sum_zm_z\phi_z}$ projects to a term $\propto\prod_z\gamma_z^{m_z}$, whose parity under $\Pi=i^{Z/2}\gamma_1\gamma_2\cdots\gamma_Z$ is $(-1)^{\sum_zm_z}$, so harmonics with $\sum_zm_z=0$ produce only even terms. The rule does not rely on conservation of $N_{\mathrm{tot}}=\sum_zN_z$, which by Eq.~\eqref{eq:Ntot_not_conserved} fails for $H_{\mathrm{linear}}$ itself. In the superconducting circuit language, the term $\sim \widetilde{h}^{(2)}$ (regular, bilinear fermion tunneling) corresponds to transporting single Cooper pairs, whereas two-body interactions $\sim\widetilde{h}^{(4)}$ correspond higher order tunneling processes (e.g., Cooper pair co-tunneling, or quartet physics).
Note further that Eq.~\eqref{eq_H_Majorana} is more general than Eq.~\eqref{eq_fermion_H}, as it also allows the inclusion of pairing, $\sim c^\dagger_z c^\dagger_{z^\prime}$. As already foreshadowed, the Majorana Hamiltonian with quartic interactions ($\sim \widetilde{h}^{(4)}$) was recently studied in connection with so-called fracton excitations~\cite{you2019building,pretko2020fracton}.
If the quartic term dominates and its elements take random values, it falls into the area of the applicability of the SYK model~\cite{Sachdev_1993,Kitaev_2015}. This suggests a route to the SYK model, provided the quartic couplings can be made dominant and random; quartic terms require four-terminal junctions.

\section{\bluehead{Breaking Wigner superselection rule}}\label{sec_fermion_parity}

The mapping also leads to field theories with unusual properties. As a first example, we keep the Chern matrix above, $\mathcal C^{(1)}_{zz'}=2$ for all pairs, so that the projected operators form a Majorana algebra. We now drop the restriction of $H_J$ to phase differences: $\epsilon_0$ and $\delta A_z$ may depend on all phases $\phi_z$ in an arbitrary but periodic way, so that Fourier components with $\sum_zm_z\neq0$ appear. Breaking the conservation of the total charge $N_{\mathrm{tot}}$ in this way is natural both in superconducting circuits, for instance through a Josephson contact to ground [Fig.~\ref{fig:ifermion}(a)], and in bosonic networks, through an ac drive. In superconducting circuits it is in fact the \textit{default scenario}, because, as elaborated above, nonzero Chern numbers require multiterminal junctions coupling to ground, making it difficult to avoid Cooper pair leakage. Charge conservation and charge quantization are distinct: a Josephson coupling to ground breaks the former and keeps the latter. In a superconducting circuit, $\sum_zm_z\neq0$ means that the nodes exchange Cooper pairs with the superconducting ground. Restoring the ground phase $\phi_g$, the corresponding term reads $e^{i\sum_zm_z(\phi_z-\phi_g)}$, which conserves the total charge of the enlarged circuit, nodes plus ground, while all phases, including $\phi_g$, remain $2\pi$-periodic. Consequently, circuits without targeted suppression of Josephson leakage to ground generically map onto Hamiltonians of the form
\begin{equation}\label{eq_H_Majorana_odd}
H=\sum_{z_{1}}\widetilde{h}^{(1)}_{z_{1}}\gamma_{z_{1}}+\sum_{z_{1}z_{2}}\widetilde{h}^{(2)}_{z_{1}z_{2}}\gamma_{z_{1}}\gamma_{z_{2}}+\sum_{z_{1}z_{2}z_{3}}\widetilde{h}^{(3)}_{z_{1}z_{2}z_{3}}\gamma_{z_{1}}\gamma_{z_{2}}\gamma_{z_{3}}+\ldots\ .
\end{equation}

A contact to ground thus provides terms that are \textit{linear} in the Majorana operators, $\sim\sum_z\widetilde h^{(1)}_z\gamma_z$ with real $\widetilde h^{(1)}_z$, and even cubic terms, $\sim\widetilde h^{(3)}$. By Eq.~\eqref{eq:ground_lemma}, such terms are generic rather than optional: a nonsingular Chern matrix requires the coupler itself to exchange Cooper pairs with ground. The linear terms are the single-node harmonics, $m_z=1$ and $m_{z'\neq z}=0$, that is, Josephson couplings of node $z$ to ground, and every harmonic with odd $\sum_zm_z$ contributes odd terms. These terms are Hermitian, yet they break the fermion-parity superselection rule~\cite{Wick_1952}, also known as Wigner superselection. On a physical level, the breaking of Wigner superselection is thus tied to the breaking of charge conservation in the circuit. Formally, odd terms can always be made even by attaching an auxiliary Majorana operator to each of them, and Hamiltonians of this extended form, such as spin chains in a longitudinal field, are well studied. Where parity-odd terms have been considered for fermionic systems, for instance in Jordan--Wigner encodings on qubit hardware, they have been treated as perturbations specific to the hardware, that is, as errors. Here they arise without any auxiliary mode and are the generic case, so that Hamiltonians of the form of Eq.~\eqref{eq_H_Majorana_odd} are the natural low-energy theory of a Chern network. Their collective behaviour, such as quantum phase transitions, has to our knowledge received little attention, and calls for field-theoretic methods for parity-breaking fermionic systems and their parafermionic generalizations.

Read as a lattice field theory, Eq.~\eqref{eq_including_capacitances} contains only local terms, yet its low-energy limit contains terms that the Wigner superselection rule forbids for elementary fermions. One must therefore ask whether this is consistent. It is, and no fundamental law is violated. The low-energy Hamiltonian, Eq.~\eqref{eq_H_low}, is built on the ground states $|0(\{\phi_z\})\rangle$ of $H_0$. In a superconducting realization, $H_0$ conserves fermion parity, as any theory of electrons must at some level. Every $|0(\{\phi_z\})\rangle$ therefore respects fermion-parity superselection for the underlying electrons, whatever the phases. The low-energy theory only produces coherent superpositions of these states with $\phi_z$-dependent amplitudes. Hence any two low-energy states, such as $|\psi\rangle$ and $c_z^\dagger|\psi\rangle$ [Eq.~\eqref{eq_gamma_vs_c}], have the same fermion parity in terms of the microscopic electrons. Fermion parity is broken only for the \textit{emergent} projected operators that satisfy Eq.~\eqref{eq_anyons}, as a consequence of the Landau-level physics on the closed torus. Theories of the form of Eq.~\eqref{eq_H_Majorana_odd} can also be engineered by other means, for instance with explicit Jordan--Wigner strings on qubits. What differs here is the way they arise: by default, from charge exchange with ground, rather than by design.

The anticommutation between different nodes is supplied by the Chern couplers. The microscopic terms $e^{i\phi_z}$ and $e^{i\phi_{z'}}$ commute, and their projections fail to commute only through virtual excitations to higher Landau levels of the network that couples the nodes. The couplers thus play the role of the Jordan--Wigner strings of qubit encodings~\cite{Whitfield_2011}, but no string of gates has to be applied. Three single-node pulses then suffice for a two-body gate between any pair of nodes: a pulse of a junction to ground at node $z$ realizes $e^{i\theta\gamma_z}$, and
\begin{equation}\label{eq:all_to_all}
e^{i\pi\gamma_z/4}\,e^{i\theta\gamma_{z'}}\,e^{-i\pi\gamma_z/4}=e^{i\theta\,(i\gamma_z\gamma_{z'})}\ .
\end{equation}
Every operator is applied locally, at a single node; the non-locality resides in the Chern couplers and in the low-energy projection. Gates between distant nodes are therefore synchronous by construction, on time scales long compared with the inverse cyclotron frequency.

The fermion-parity breaking found here is therefore conceptually similar to implementing fermionic operators on universal qubit hardware. On qubits, a local gate cannot act as a bare Majorana operator on a distant site; this requires a Jordan--Wigner string of nonlocal gates~\cite{Whitfield_2011}. Here each such operator is realized by a single local junction, because the string is built into the Chern couplers. This difference between explicit and built-in nonlocality is the conceptual point of this section.

The point matters for a research area that derives the axioms of quantum mechanics from information-theoretic principles rather than physical ones~\cite{Deutsch_2000,Clifton_2003,Peres_2004,Bub_2005,Chiribella_2011,Raymond-Robichaud_2021}. Within it, it is debated whether Wigner superselection for fermions requires relativistic arguments, or whether it follows in a non-relativistic setting from information-theoretic axioms such as local realism in the quantum-mechanical sense~\cite{Friis_2013,Friis_2016,johansson2016comment,Vidal_2022}.

Our theory provides a counter-example to the non-relativistic argument. It breaks Wigner superselection for the encoded fermions and therefore allows protocols that evade the no-communication argument for local fermionic modes~\cite{johansson2016comment}. Non-local couplings alone do not have this effect: a dense inverse capacitance matrix, or a non-local Coulomb interaction, couples all nodes without breaking local realism. What matters are parity-odd Hamiltonians of the form of Eq.~\eqref{eq_H_Majorana_odd}, which are generic here. Relativity is not violated, because the theory is a low-energy theory. The apparent instantaneity holds only within the Born--Oppenheimer limit: the electromagnetic fields and the coupler have been integrated out, and operations faster than the response of the coupler, set by $\Delta_{\min}$, leave the regime of validity. Imposing no-signalling without relativistic arguments~\cite{johansson2016comment} would nonetheless exclude a theory like the present one from the outset. Open questions on the collective properties of such Hamiltonians are collected in Sec.~\ref{sec_conclusions_outlook}.

In Appendix~\ref{app:protocol_nocommun} we give a circuit realization of a simple protocol from Ref.~\cite{johansson2016comment}. It shows how breaking Wigner superselection lets local gates on Bob's nodes change the outcome of a measurement on Alice's nodes. Alice and Bob each act only on their own nodes, through junctions to ground and gate charges; the non-locality is embedded in the Chern couplers, not in the operations. The protocol needs only two fermionic modes, one for Alice and one for Bob, and therefore runs on a circuit with four Majorana nodes.

The protocol thus serves two purposes. It tests the encoded anticommutation directly, and it probes the no-communication argument in this setting, with the all-to-all gates of Eq.~\eqref{eq:all_to_all} as the practical counterpart. It does not require precise gate operations or strong error suppression, and it is sensitive to weak parity-odd terms.
\section{\bluehead{Non-identical exchange algebras}}\label{sec_nonidentical}

The connection between locality and charge conservation established above holds when all the Chern numbers in a complete graph are equal, $\mathcal C_{zz'}=\mathcal C$. For $\mathcal C=2$ the projected operators are Majorana fermions, and the fermionic Wigner superselection rule is recovered for Hamiltonians built from phase differences. The superselection principle generalizes to parafermions for $\mathcal C>2$: the operators $\eta_z\equiv e^{i\Phi_z}$ satisfy $\eta_z^{\mathcal C}=1$ and $\eta_z\eta_{z'}=e^{i2\pi/\mathcal C}\eta_{z'}\eta_z$ for $z<z'$. For $\mathcal C>2$ this requires the couplers to alternate in orientation, $\mathcal C^{(1)}_{zz'}=(-1)^{z+z'+1}\mathcal C$ for $z<z'$, so that all phases $q_{zz'}/p$ are equal; for identical orientation they alternate in sign. With this condition, the exchange phases do not depend on $z$ and $z'$, a property we call an \textit{identical} exchange algebra. Imposing the same charge-conservation constraint on the full $H$, projection onto the lowest Landau level then gives a Hamiltonian of the form
The orientation condition has a simple graphical meaning. Draw each coupler as an arrow from $z$ to $z'$ when $\mathcal C^{(1)}_{zz'}>0$. Identical orientation turns the complete graph into a transitive tournament, which has no directed three-cycle, whereas alternating orientation contains directed three-cycles, such as $1\to2\to3\to1$. Whether a tournament is transitive does not depend on how the nodes are labelled, so no permutation of the nodes maps one orientation onto the other. The relabelling $\phi_z\to(-1)^z\phi_z$ does, up to reversing all arrows, but it maps neutral tunnelling terms onto terms that exchange charge with ground.
\begin{equation}\label{eq_H_parafermion_quadratic}
    H=\sum_{z_1 z_2}h_{z_1,z_2}\eta_{z_1}^\dagger \eta_{z_2}+\sum_{z_1 z_2 z_3 z_4}h_{z_1 z_2 z_3 z_4}^{(4)}\eta_{z_1}^\dagger \eta_{z_2}^\dagger \eta_{z_3}\eta_{z_4}+\ldots\ .
\end{equation}
The generalized parity operator~\cite{Fendley_2012} (also called symmetry generator) can be defined as
\begin{equation}\label{eq:generalized_parity}
    Q=\prod_{z=1}^{Z/2}\eta_{2z-1}^\dagger \eta_{2z}\ ,
\end{equation}

The operator $Q$ defined in Eq.~\eqref{eq:generalized_parity} is unitary and satisfies $Q^\dagger\eta_zQ=e^{i2\pi/\mathcal C}\eta_z$ for all $z$. The charge-conserving Hamiltonian above therefore obeys $Q^\dagger HQ=H$, so $Q$ is a conserved generalized parity. This leads again to a no-signalling theorem, provided no unitary operation or measurement by two participants acting on different nodes, Alice and Bob, can create superpositions of states with different eigenvalues of $Q$. On the present platform, with operations and measurements restricted to charge-conserving processes and, optionally, gates that shift charge and flux, this generalized Wigner superselection anchored to $Q$ holds for every communication experiment the hardware can run. If charge conservation is relaxed again, which adds non-bilinear terms to $H$ (Sec.~\ref{sec_fermion_parity}), the no-signalling theorem can again be broken.

Crucially, no-signalling breaks down even with charge conservation if the Chern numbers are unequal. The projected operators $\eta_z=e^{i\Phi_z}$ then obey the exchange algebra of Eq.~\eqref{eq_anyons} with phases $q_{zz'}/p$ that depend on $z$ and $z'$; we call such an algebra \textit{non-identical}. By Eq.~\eqref{eq:frobenius}, the underlying algebra is always a product of independent clock--shift pairs in a rotated integer basis; what the pair dependence of $q_{zz'}$ encodes is which local Josephson terms act as which of their Weyl operators. A charge-conserving theory still maps onto a Hamiltonian of the form of Eq.~\eqref{eq_H_parafermion_quadratic}, but this theory is no longer local, as we now show in the language of quantum communication, independently of a specific Hamiltonian.

We partition the $Z$ nodes into two disjoint sets, one for Alice, $z=1,\ldots,Z_A$, and one for Bob, $z=Z-Z_B+1,\ldots,Z$, with $Z=Z_A+Z_B$. Alice and Bob may perform unitary operations $U_A=U_0e^{iA}$ and $U_B=V_0e^{iB}$, with Hermitian $A$ and $B$ and unitary $U_0$ and $V_0$, or measure Hermitian operators $M_A$ and $M_B$. We restrict Alice and Bob to neutral operations, built from Josephson harmonics with $\sum_zm_z=0$, together with charge and flux shifts. Because $H_{\mathrm{linear}}$ does not conserve $N_{\mathrm{tot}}$, Eq.~\eqref{eq:Ntot_not_conserved}, we impose this restriction on the harmonics directly rather than through a conservation law. On the torus, the symmetry that survives is the $\mathbb Z_g$ remnant $T=W(\mathcal C^{(1)}\mathbf 1/g)$ of the total charge, where $g$ is the greatest common divisor of the row sums of $\mathcal C^{(1)}$: $T$ commutes with $W(\mathbf n)$ if and only if $\sum_zn_z\equiv0$ modulo $g$, and for the Majorana network $T$ is the fermion parity. The restriction acts as a charge superselection rule: neither unitaries nor projective measurements may create superpositions of states with different total charge. For the fermionic case this is the Wigner superselection rule. Projected onto the lowest Landau level, the rule restricts $A$, $B$, $M_A$ and $M_B$ to the same expansion as the Hamiltonian in Eq.~\eqref{eq_H_parafermion_quadratic}, so that $e^{iA}$ and $e^{iB}$ cannot change the total charge. This is too restrictive for unitaries, because superselection forbids only superpositions of different charges, not changes of charge. The prefactor unitaries $U_0=\prod_ze^{i\alpha_z\Phi_z}$ and $V_0=\prod_ze^{i\beta_z\Phi_z}$, with integer $\alpha_z$ and $\beta_z$, account for such changes. For identical exchange algebras, $q_{zz'}=q$ for all $z>z'$, locality, meaning that $U_A$ and $M_A$ contain only Alice's operators $\eta_z$ and $U_B$ and $M_B$ only Bob's, guarantees that Alice's operations commute with Bob's. For non-identical exchange algebras, with $q_{zz'}$ depending on $z$ and $z'$, they need not commute, although Alice and Bob act only on their own nodes. As in Sec.~\ref{sec_fermion_parity}, the non-commutation is supplied by the Chern couplers. This breaks locality in node space, and with it common notions of locality.

The criterion follows directly from Eq.~\eqref{eq_anyons}. With $\Theta=(\mathcal C^{(1)})^{-1}$, neutral harmonics on disjoint node sets $A$ and $B$ are generated by tunnelling vectors $\mathbf e_a-\mathbf e_{a'}$ and $\mathbf e_b-\mathbf e_{b'}$. Their exchange phase is $e^{-2\pi i(\Theta_{ab}-\Theta_{a'b}-\Theta_{ab'}+\Theta_{a'b'})}$, so all neutral processes commute precisely when the combination in parentheses is an integer for every $a,a'\in A$ and $b,b'\in B$. In the chain of Sec.~\ref{sec_LL_torus}, tunnelling between nodes 1 and 2 and tunnelling between nodes 3 and 4 acquire a factor $-i$ on exchange, although no coupler joins the two pairs.

We thus have two versions of locality breaking: identical exchange algebras combined with a breaking of charge superselection, defined on the original boson field; or non-identical exchange algebras, for which locality breaks whether or not charge superselection holds. The first can be treated on the same footing as the second. Charge conservation is broken by couplings to ground, which arise from the Fourier components of $\epsilon_0$ and $\delta A_z$ with $\sum_zm_z\neq0$, see Eq.~\eqref{equation_epsilon_Fourier}. Ground can be included explicitly through the global shift $\phi_z\to\phi_z-\phi_g$, the usual $U(1)$ gauge of charge conservation. Giving ground a finite, though very large, capacitance in Eq.~\eqref{eq_including_capacitances} promotes $e^{i\phi_g}$ from a scalar to an operator that counts the charges exchanged with ground. For identical exchange algebras, $\mathcal C_{zz'}=\mathcal C$, every unpaired Majorana or parafermion operator [for Majoranas, the odd terms in Eq.~\eqref{eq_H_Majorana_odd}] then acquires a counting-field operator, $\eta_z\to\eta_ze^{-i\phi_g}$, which tracks the total charge including ground. Since no Chern connections (gyrators) link the nodes to ground, $e^{i\phi_g}$ commutes with all $\eta_z$ and cannot salvage locality. The ground phase operator thus acts as an additional node whose exchange relations differ from the fermionic or parafermionic ones, a special case of a non-identical exchange algebra.

\section{\bluehead{Conclusions and outlook}}\label{sec_conclusions_outlook}

\subsection{\bluehead{Conclusions}}

In this work, we demonstrated that geometric (Berry curvature) terms in suitable conventional superconducting multiterminal junctions, namely those with a flat ground band, give rise to projected phase operators with a rational exchange algebra. The fractional exchange phase $\sim q/p$ is linked to the nonzero Chern numbers of the system, and can take in principle any rational value, such that there emerge Majoranas or even more general types of parafermions. The fractional exchange phases can be measured by a flux period reduced to $\Phi_0/p$ and by the fractional shift that an offset charge induces in the flux dependence. Per branch, this is a fractional dual Josephson effect, the charge counterpart of the $4\pi$-periodic Josephson effect of Majorana junctions, protected by a conserved $\mathbb Z_p$ label rather than by fermion parity. Due to the emergent exchange algebra, the here considered system allows direct implementation of a simulation of fermionic systems with \textit{local} gate operations, relevant for a number of important problems, such as quantum computational chemistry.

Four structural results sharpen this picture. A nonsingular Chern matrix requires the coupler to exchange Cooper pairs with ground, Eq.~\eqref{eq:ground_lemma}, which makes terms with $\sum_zm_z\neq0$ generic. Every Landau level carries a single $|\mathrm{Pf}\,\mathcal C|$-dimensional representation of a product of clock--shift pairs, Eq.~\eqref{eq:frobenius}. The controlled regime requires a coupler with a flat ground band, Eq.~\eqref{eq:flatband}, an engineering condition independent of topology. Finally, a flux-biased junction at node $z$ has flux period $\Phi_0/g_z$, a direct spectroscopic signature of the Chern matrix.

Our work also provides a number of insights of fundamental nature. We particularly highlighted the fact that there naturally occur interactions with \textit{non-identical} exchange algebras, ultimately yielding a lattice quantum theory that breaks fermion-parity superselection rule, and generalizations thereof for parafermions. This effect can be related on a circuit level to the breaking of charge conservation with respect to ground. 
In contrast to other known parity breaking effects (as in implementations of fermionic gates in qubit hardware via non-local Jordan-Wigner chains), the local nature of the fermionic gates allowed us to connect our work to the ongoing endeavour to derive quantum mechanics from an information-theoretic point of view, suggesting that relativistic arguments are needed to enforce the no-communication constraint.

\subsection{\bluehead{Outlook}}

By including geometric interactions in generic many-body lattice systems, we lay the groundwork towards a different flavour of quantum information hardware, where a qudit is directly encoded in a magnetic-translation representation with a pair-dependent, \textit{non-local} exchange behaviour.
Beyond applications, the here developed theory presents a whole host of important open questions of fundamental nature. For instance, the aforementioned breaking of fermion-parity superselection rules (and parafermionic generalizations thereof) stipulates that the understanding of networks with quantum geometric interactions involves solving a class of quantum field theories which have so far been treated mainly as hardware errors rather than as model Hamiltonians. Take for concreteness the ordinary Hubbard model, which can be realized in circuits with charge-conserving Josephson effects, see Eq.~\eqref{eq_fermion_H}. We would for instance deem it highly interesting to investigate if the known phenomenology, such as the phase diagram (for which, incidentally, there still does not exist full consensus~\cite{Arovas_2022_review}), survives the addition of (even weak) fermion-parity breaking processes. Whether parity-odd terms change ground-state entanglement is also a computational question. Entanglement scaling bears on the accuracy of matrix-product-state approximations and the cost of density-matrix renormalization group calculations~\cite{Schollwoeck_2005_review,Eisert_2010_review,Verstraete_2006}.

Our setup shares with chiral-edge theories of the fractional quantum Hall effect the first-order structure of the phase dynamics, but not their braiding. It realizes a reduced (i.e., scalar) version of Chern-Simons theory, in the sense of the topological quantum mechanics of Ref.~\cite{Dunne_1990}, with a number of special properties, most notably the absence of a bulk in the proper sense (the Chern number is defined within the node degrees of freedom itself), and the possibility of a spatially varying Chern level, which is at the origin of the Wigner superselection breaking. In an upcoming work, we plan to deepen the understanding of the change of the usual bulk--boundary correspondence found here, which we will explore as \textit{hollow} topological matter.
Finally, we assumed throughout a nonsingular Chern matrix. If $\det\mathcal C^{(1)}=0$, for instance for an odd number of nodes or for a coupler without a terminal at ground, Eq.~\eqref{eq:ground_lemma}, the combinations of phases in the kernel of $\mathcal C^{(1)}$ are not affected by the Chern couplings and remain ordinary rotor degrees of freedom, while the Landau-level construction applies to the complementary combinations, with kinetic flux operators built from the nonzero eigenvalues of $g-g^T$ only.

\section*{\bluehead{Acknowledgements}}
We warmly thank Fabian Hassler, David DiVincenzo, Markus Müller, Maarten R. Wegewijs, Frank Wilhelm-Mauch, Z. Leghtas, P. Campagne-Ibarcq, J.-D. Pillet, L. Bretheau, and Janine Splettstoesser for fruitful discussions. R.M. would like to thank Julia Meyer for her insights and for directing us to relevant literature.
The scientific results, theoretical developments, and derivations presented in this work were completed in 2024, and the first version of the manuscript was prepared without the use of generative AI. During the preparation of the present revised version, the authors used OpenAI's GPT-5.6 Sol and Codex solely to expedite language editing and improve the clarity, readability, and presentation of selected passages. The scientific content and conclusions remain the authors' own, and all authors take full responsibility for the final manuscript.
\paragraph{Funding information}
This work has been funded by the German Federal Ministry of Education and Research within the funding program Photonic Research Germany under the contract number 13N14891.

\appendix
\section{\bluehead{Cavity Chern insulator with charge qubit coupler}}\label{app:cavity_supp}

The standard Chern insulator Hamiltonian in a 2D lattice is given as
\begin{equation}\label{eq_CI_target}
    H_\text{CI}=\left[\delta+\cos\left(\phi_{1}\right)+\cos\left(\phi_{2}\right)\right]\sigma_{z}+\sin\left(\phi_{1}\right)\sigma_{x}+\sin\left(\phi_{2}\right)\sigma_{y}\ .
\end{equation}
For simplicity, above Hamiltonian is unitless. We are free to multiply the entirety of $H_\text{CI}$ with a total energy scale $h$, i.e., $H_\text{CI}\rightarrow h H_\text{CI}$. This Hamiltonian is well-known to exhibit topological gapped phases for $\vert\delta\vert<2$, with Chern number $\pm 1$ (defined in $\phi_1$ and $\phi_2$). Its ground band is not flat. With $\mathbf v=(\sin\phi_1,\sin\phi_2,\delta+\cos\phi_1+\cos\phi_2)$, one has $\min|\mathbf v|=\min(|\delta|,2-|\delta|)$ and $\max|\mathbf v|=2+|\delta|$, so Eq.~\eqref{eq:flatband} gives $\Delta\epsilon_0/\Delta_{\min}\geq1$ throughout the topological phase, with equality at $|\delta|=1$. The model thus demonstrates a Chern-nontrivial coupler, not the hierarchy $\Delta\epsilon_0\ll\omega\ll\Delta_{\min}$. The flattened Hamiltonian $h\,\mathbf v\cdot\boldsymbol\sigma/|\mathbf v|$ has the same Chern number and $\Delta\epsilon_0=0$, but is no longer of the simple form of Eq.~\eqref{eq_H_cavity}.

Let the above be our target Hamiltonian that we seek to implement with the photonic degrees of freedom of a quantum optical cavity. Specifically, if we use two cavity modes $b_1^{(\dagger)}$ and $b_2^{(\dagger)}$, see also right panel in Fig.~\ref{fig:qcgyrator}c, we can reverse engineer above Chern insulator Hamiltonian via the bosonic cavity Hamiltonian
\begin{equation}\label{eq_H_cavity}
\begin{split}
H_{\text{cavity}}=\widetilde{\delta}\sigma_{z}+\frac{b_{1}+b_{1}^{\dagger}}{2}\sigma_{z}+\frac{b_{2}+b_{2}^{\dagger}}{2}\sigma_{z}+i\frac{b_{1}-b_{1}^{\dagger}}{2}\sigma_{x}+i\frac{b_{2}-b_{2}^{\dagger}}{2}\sigma_{y} \\+u_{K1}\left(b_{1}^{\dagger}b_{1}-\mu_{1}\right)^{2}+u_{K2}\left(b_{2}^{\dagger}b_{2}-\mu_{2}\right)^{2}+u_{cK}b_{1}^{\dagger}b_{1}b_{2}^{\dagger}b_{2}\ .
\end{split}
\end{equation}
The second line contains the free boson Hamiltonian with auto- and cross-Kerr terms ($u_{K1}$ and $u_{K2}$, respectively, $u_{cK}$) representing the boson many-body interactions within the cavities (auto-Kerr) and between the two cavities (cross-Kerr). The offsets (chemical potentials) $\mu_{1,2}$ can be suitably tuned by an ac drive to the individual cavities. The logic is now as follows. By tuning $\mu_{1,2}$ to large values, we can drive the cavities to a high-photon population state, $n_{1,2}=b_{1,2}^\dagger b_{1,2}\gg 1$. In that regime, the boson operators can be approximated as $b_{1,2}\approx\sqrt{n_{1,2}}e^{-i\phi_{1,2}}$, thus mapping from Eq.~\eqref{eq_H_cavity} to Eq.~\eqref{eq_CI_target}. Note that in these two Hamiltonians, the amplitude of the $\sigma_z$ term is rescaled ($\delta$ versus $\widetilde{\delta}$). This rescaling is necessary to account for the population prefactor $\sim \sqrt{n_{1,2}}$ when replacing the boson ladder operators with the phase operators.

The remaining challenge now is to show that a suitably designed capacitive and inductive coupling between a given cavity mode and, e.g., a charge qubit can implement all necessary interaction terms, given in the first line of Eq.~\eqref{eq_H_cavity}. We consider an ordinary flux-tunable charge qubit, with a dc-SQUID composed of two Josephson junctions, shunted by a capacitance, as shown in Fig.~\ref{fig:qcgyrator}c. Let us for simplicity start by considering a single cavity with modes $X=(b+b^\dagger)/\sqrt{2}$ and $P=i(b-b^\dagger)/\sqrt{2}$, representing the electro-magnetic charge and flux of the cavity, respectively. A generic capacitive and inductive coupling between cavity and charge qubit yields the Hamiltonian
\begin{equation}
    H_{cc}=E_{C}\left(n+n_{\text{ext}}+\frac{\mu}{E_{C}}\widehat{X}\right)^{2}-\left(E_{J}+2\nu\widehat{P}\right)\cos\left(\varphi+2\frac{\lambda}{E_{J}}\widehat{P}\right)\ .
\end{equation}
The charge qubit (with charge and phase $[\varphi,n]=i$) is modelled with the following parameters: the capacitive energy $E_C$ and the total Josephson energy $E_J$ of the SQUID (which is tunable by an externally applied flux). The capacitive (charge-charge) coupling is parametrized by the energy scale $\mu$. The inductive (current-current) coupling enters twice: once in an offset of the Josephson cosine energy (with the energy scale $\lambda$), and a second time as a dynamical rescaling the Josephson energy itself (proportional to $\nu$). We tune the offset charge $n_\text{ext}$ close to the charge degeneracy point, and operate the charge qubit close to the Cooper pair box regime ($E_C > E_J$). For instance, for $n_\text{ext}=-1/2+\delta n_\text{ext}$ ($\delta n_\text{ext}\ll 1$), the two (nearly) degenerate charge states are $n=0,1$, forming the charge qubit basis with the Pauli operators $n-1/2\rightarrow \sigma_z/2$, $\cos(\phi)\rightarrow \sigma_x/2$, and $\sin(\phi)\rightarrow \sigma_y/2$. Projected onto that qubit basis (and up to linear order in all couplings $\mu,\nu,\lambda$), we get
\begin{equation}
    H_{cc}\approx\widetilde{\delta}\sigma_{z}-\widetilde{\gamma}\sigma_{x}+\mu\sigma_{z}X-\nu \sigma_{x}P+\lambda\sigma_{y}P\ ,
\end{equation}
where $\widetilde{\delta}=E_C\delta n_\text{ext}$ and $\widetilde{\gamma}=E_J/2$. Tuning sufficiently far off resonance, such that $\widetilde{\delta}$ is larger than $\widetilde{\gamma}$, we are almost at the desired shape. Now, we include two cavities instead of one, $X,P\rightarrow X_{1,2},P_{1,2}$. In lowest order in the coupling, the interactions between charge qubit and the two cavities is simply additive, yielding\begin{equation}
    H_{cc}\approx\widetilde{\delta}\sigma_{z}+\sum_{j=1,2}\left[\mu_j\sigma_{z}X_j-\nu_j \sigma_{x}P_j+\lambda_j\sigma_{y}P_j\right]\ .
\end{equation}
All that remains is to tune the couplings to values $\mu_1\approx \nu_1 \approx \mu_2 \approx \lambda_2 \approx 1$ and $\lambda_1\approx \nu_2\approx 0$ to arrive at the target Hamiltonian given in Eq.~\eqref{eq_H_cavity}. This concludes the engineering of a Chern insulator in the 2D lattice space defined by the photon numbers of two cavities.

\section{\bluehead{Gauges for a uniform magnetic field on the torus}}\label{app:gauges_supp}

As noted in the main text, usual gauge choices for vector potentials in the extended 2D plane are at risk of yielding inconsistent results when embedded on a compact manifold, such as the torus. We here clarify certain gauge aspects for said torus, by means of a numerical investigation.

The linear vector potential $a_z=\sum_{z'}g_{zz'}\phi_{z'}$ of Eq.~\eqref{eq_A_g} is not $2\pi$-periodic and is therefore not a valid gauge on the torus as it stands. A consistent gauge follows from a curvature $\widetilde{\mathcal B}_{zz'}$ that differs from the Berry curvature by $\mathcal C^{(1)}_{zz'}$ Dirac-delta ``horns''~\cite{Onofri_2001},
\begin{equation}\label{eq:horns_app}
    \widetilde{\mathcal{B}}_{zz^\prime}=\mathcal{B}_{zz^\prime}-2\pi\sum_{i}\mathcal{K}_{zz^\prime}^{(i)}\delta(\phi_z-\phi_z^{(i)})\,\delta(\phi_{z^\prime}-\phi_{z^\prime}^{(i)})\ ,
\end{equation}
with integer heights $\mathcal K^{(i)}_{zz'}$ at arbitrary positions $(\phi_z^{(i)},\phi_{z'}^{(i)})$ and $\sum_i\mathcal K^{(i)}_{zz'}=\mathcal C^{(1)}_{zz'}$. By construction, the integral of $\widetilde{\mathcal B}$ over each two-torus vanishes,
\begin{equation}\label{eq:gaussbonnet_app}
    \frac{1}{2\pi}\int_0^{2\pi}d\phi_z\int_0^{2\pi}d\phi_{z'}\,\widetilde{\mathcal B}_{zz'}=0\ ,
\end{equation}
as required by Stokes' theorem for the curl of a periodic vector potential on a closed surface. $\widetilde{\mathcal B}$ is therefore the curl of a vector potential that is locally given by Eq.~\eqref{eq_A_g} but contains jumps that restore $2\pi$-periodicity. The horns do not affect the dynamics, because the phase acquired by encircling one is $e^{i2\pi}=1$. One may therefore work with $\mathcal B$ in any representation whose operator algebra respects phase compactness, such as the projected operators of Sec.~\ref{sec_LL_torus}. The construction is convenient for numerics on a discretized torus, which we now describe for $Z=2$.

We discretize the torus with a mesh of $N\times N$ points, as shown in Fig.~\ref{fig:discretizing}.
The Hamiltonian resulting from Eq.~\eqref{eq_L_gyrator} (under appropriate inclusion of capacitive coupling) can be numerically implemented as the single-particle Hamiltonian hopping between the points on the mesh, and the effective magnetic field can be implemented as the phase acquired along the movement between the neighbour points.

Thus, the Berry connection in this discretized model is described by the values $A^{x}_{n,m}= \int_{x_{n}}^{x_{n+1}} A_{x}(x,y_{m}) dx$ and same formula for $A^{y}_{n,m}$ with $x\leftrightarrow y$, and fluxes through the unit cell $\tilde{\mathcal{B}}_{nm}= \int_{x_{n}}^{x_{n+1}}\int_{y_{m}}^{y_{m+1}} \tilde{\mathcal{B}}(x,y) dxdy$.
Here, to simplify the notation, we denote $x=\phi_{1}$, $y=\phi_{2}$. Periodic boundary conditions are implemented by using cyclic indices $n,m$, that is, they are set to $0$ when they assume the value $N$.
The constant magnetic field $\mathcal{B}$ is implemented through the condition
\begin{gather}
\tilde{\mathcal{B}}_{nm} = A^{x}_{n,m}+A^{y}_{n+1,m}-A^{x}_{n,m+1}-A^{y}_{n,m},\\
\exp\left\{i\tilde{\mathcal{B}}_{nm}\right\} = \exp\left\{i(2\pi/N)^{2}\mathcal{B}\right\},
\end{gather}
where $\tilde{\mathcal{B}}_{nm}$ is basically the flux through the elementary cell.
The exponentiation of the phases relates the values up to the $2\pi$, giving us the relation
\begin{equation}
\tilde{\mathcal{B}}_{nm} = (2\pi/N)^{2}\mathcal{B} - 2\pi \mathcal{K}_{nm},\qquad \mathcal{K}_{nm} \in \mathbb{Z}.
\end{equation}
On the other side, using the definition of $\tilde{\mathcal{B}}_{nm}$ and cyclic indices, we get $\sum_{\forall n,m}\tilde{\mathcal{B}}_{nm} = 0$, which naturally follows from Stokes theorem on the torus.
This condition results in the quantization of the magnetic field
\begin{equation}
2\pi \mathcal{B} = \sum_{\forall n,m} \mathcal{K}_{nm} = \mathcal{C} \in \mathbb{Z},
\end{equation}
where $\mathcal{C}$ is the Chern number. Here, the most trivial example is when all ``horns'' $\mathcal{K}$ [see Eq.~\eqref{eq:horns_app} are gathered into a single one at $(n_0,m_0)$, i.e.\ $\mathcal{K}_{nm}=0$ everywhere except for $\mathcal{K}_{n_{0}m_{0}}=\mathcal{C}$.
Then, for example a linear gauge can be constructed in the following way:
\begin{equation}
\begin{aligned}
\!\!\!A^{x}_{n_{0},m\leq m_{0}} & \!=\! -\frac{2\pi\mathcal{C}}{N}m, &
\!A^{x}_{n_{0},m>m_{0}} & \!=\! -\frac{2\pi\mathcal{C}}{N}m \!+\! 2\pi\mathcal{C},
\\
\!\!\!A^{y}_{n \leq n_{0},m} & \!=\! \frac{2\pi\mathcal{C}}{N^{2}}n, &
\!A^{y}_{n > n_{0},m} & \!=\! \frac{2\pi\mathcal{C}}{N^{2}}n - \frac{2\pi\mathcal{C}}{N}
\end{aligned}
\end{equation}
which is also illustrated in Fig.~\ref{fig:gauge}.

\begin{figure}
\centering
\begin{picture}(220,150)(-80,-60)
\multiput(-20,-20)(15, 0){7}{\multiput(0 ,0)(0, 15){7}{\circle{3}}}
\multiput( 10, 10)(15, 0){4}{\multiput(0 ,0)(0, 15){4}{\circle*{3}}}
\multiput(-5, 0)(45, 0){2}{\multiput(0 ,-26)(0,5){21}{\line(0,1){2}}}
\multiput( 0,-5)( 0,45){2}{\multiput(-26, 0)(5,0){21}{\line(1,0){2}}}
\put(27,10){\tiny $A^{1/x}_{21}$}
\put(6,30){\tiny $A^{2/y}_{12}$}
\put(-5,-5){\vector(0,1){15}}
\put(10,-5){\vector(0,1){15}}
\put(-5,-5){\vector(1,0){15}}
\put(-5,10){\vector(1,0){15}}
\put(-30,-30){\vector(0,1){110}}
\put(-25,80){$\phi_{2}/y$}
\put(-29,52){$N$}
\put(-29, 7){$1$}
\put(-29,-8){$0$}
\put(-30,-30){\vector(1,0){110}}
\put(80,-25){$\phi_{1}/x$}
\put(52,-29){$N$}
\put( 8,-29){$1$}
\put(-7,-29){$0$}
\put(-90, -50){$H = e^{iA_{nm}^{x}}|n+1,m\rangle\langle n,m|+e^{iA_{nm}^{y}}|n,m+1\rangle\langle n,m|+h.c.$}
\end{picture}
\caption{Discretization of the phase torus of two superconducting islands. The mesh is defined on a torus, i.e.\ index $N\equiv 0$.}
\label{fig:discretizing}
\end{figure}

This discrete gauge in the continuous limit takes the following form
\begin{equation}
\begin{aligned}
A^{x}(x,y) &= - \mathcal{C} \delta(x-x_{0}) \bigl(y-2\pi\theta(y-y_{0})\bigr),
\\
A^{y}(x,y) &= \frac{\mathcal{C}}{2\pi} \bigl(x - 2\pi\theta(x-x_{0})\bigr),
\\
\widetilde{\mathcal{B}}(x,y) &= \frac{\mathcal{C}}{2\pi}
- 2\pi \mathcal{C} \delta(x-x_{0}) \delta(y-y_{0}),
\end{aligned}
\end{equation}
where $\delta(x)=\theta'(x)$ denotes a $2\pi$ periodic Dirac delta function, and the $\theta(x)$ is an extended step function, i.e., not a single step, but a ladder, with steps at $-2\pi$, $0$, $2\pi$, $4\pi$, and so on. In a gauge of this type the operators $e^{i\Phi_z}$ are manifestly $2\pi$-periodic, see Eq.~\eqref{eq:landau_gauge_translations}.

Numerical diagonalization of the Hamiltonian given in Fig.~\ref{fig:discretizing} for the case $\mathcal{C}=2$ illustrates the predicted $\mathcal{C}$-fold degeneracy of the lowest Landau level on the torus in the main text, see Fig.~\ref{fig:gauge}(b), left.
On the right-hand side of Fig.~\ref{fig:gauge}(b), the probability distribution of the wave function is shown. Here one may notice an important feature. Despite the problem being translational invariant, the physical part of the solution, which is the probability, is not. The wave function has a maximum, and its position is not related to the ``horn'', which proves that the ``horn'' does not affect the measurable part of the solution. But the phase of the wave function is, of course, affected by the gauge.
This issue arises from the fact that the gauge invariance means that all fluxes remain the same. In the case of the torus it adds two different types of closed trajectories, for example the loops $\mathcal{W}_{z}$ and $\mathcal{W}_{z'}$ shown in Fig.~\ref{fig:torus}(b) [in our case $z=1$ and $z'=2$].
Thus, we need to fix the values for the fluxes through these loops too.
Modifying the gauge, we found the correlation between these fluxes and the solution: the maximum of the wave function coincides with the crossing of the straight trajectories $\mathcal{W}_{1}$ and $\mathcal{W}_{2}$, position of which chosen such that the fluxes through these both trajectories are equal to zero (up to $2\pi$ addition, of course), see the grey lines in Fig.~\ref{fig:gauge}(b), right.

\begin{figure}[t]
\centering
\includegraphics[width=0.6\textwidth]{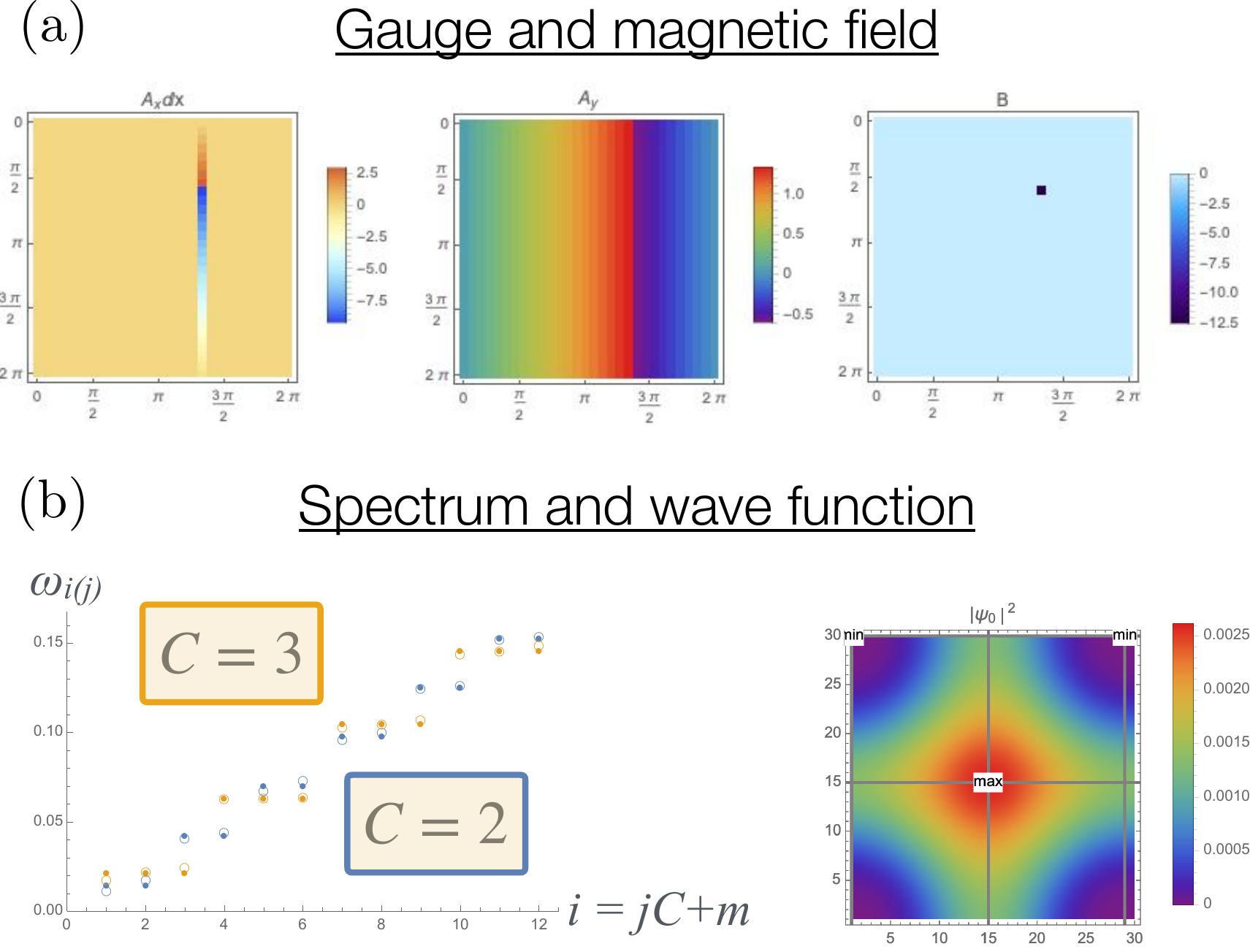}
\caption{(a) An example of the linear gauge. In the left plot the phase is scaled by $dx=2\pi/N$. In the right plot one can clearly see the ``horn'' in the magnetic field.
(b) The numerical solution for the Hamiltonian given in the Fig.~\ref{fig:discretizing}. On the left is the spectrum with a distinct degeneracy for each Landau level.
The blue and yellow circles (solid) represent the cases of the degenerate Landau levels for $\mathcal C=2$ and $\mathcal C=3$, respectively. When adding Josephson junctions, the degeneracy is lifted (empty circles).
On the right is the wave function probability. Grey lines correspond to the topologically non-trivial straight loops with zero flux through them.}
\label{fig:gauge}
\end{figure}

\section{\bluehead{Pfaffian formulas and the four-node case}}\label{app:pfaffian}

For even $Z$, the denominator $p$ in Eq.~\eqref{eq_exchange_theta} is the $Z/2$-th Chern number,
\begin{equation}\label{eq_p_app}
    p=\mathcal{C}^{(Z/2)}=\sum_{z_1,\ldots,z_Z}\frac{\varepsilon^{z_1\ldots z_Z}\,\mathcal{C}^{(1)}_{z_1z_2}\cdots \mathcal{C}^{(1)}_{z_{Z-1}z_Z}}{2^{Z/2}(Z/2)!}=\mathrm{Pf}\,\mathcal C^{(1)}\ ,
\end{equation}
where $\varepsilon$ is the Levi-Civita symbol. The integer matrix $q$ is a ``transposed'' $(Z/2-1)$-th Chern number,
\begin{equation}\label{eq_q_matrix_app}
    q_{zz^\prime}=\overline{\mathcal{C}}_{zz^\prime}^{(Z/2-1)}=\sum_{z_3,\ldots,z_Z}\frac{\varepsilon^{z^\prime z z_3\ldots z_Z}\,\mathcal{C}^{(1)}_{z_3z_4}\cdots \mathcal{C}^{(1)}_{z_{Z-1}z_Z}}{2^{Z/2-1}(Z/2-1)!}\ .
\end{equation}
It plays the role for skew-symmetric matrices that the adjugate plays for general matrices: its elements are signed Pfaffian minors,
\begin{equation}\label{eq_q_minor_app}
    q_{zz'}=(-1)^{z+z'}\,\mathrm{Pf}\,\mathcal C^{\widehat z\widehat z'}\qquad (z<z')\ ,\qquad q_{z'z}=-q_{zz'}\ ,
\end{equation}
where $\mathcal C^{\widehat z\widehat z'}$ is obtained from $\mathcal C\equiv\mathcal C^{(1)}$ by deleting rows and columns $z$ and $z'$. The Pfaffian obeys the recursion~\cite[Eq.~(2.4)]{Okada_2019}
\begin{equation}\label{eq_pf_recursion_app}
    \mathrm{Pf}\,\mathcal C=\sum_{z'=2}^{Z}(-1)^{z'}\,\mathcal C_{1z'}\,\mathrm{Pf}\,\mathcal C^{\widehat 1\widehat z'}\ .
\end{equation}
Since $\mathcal C$ is integer valued, $p$ and every element of $q$ are integers.

For four nodes, $q/p$ and the second Chern number $\mathcal C^{(2)}$ are given in Eq.~\eqref{eq_q/p_4}.
With $C_{zz'}=C_g\delta_{zz'}$, the two cyclotron modes, $H=\omega_{+}\hat a_+^\dagger \hat a_++\omega_{-}\hat a_-^\dagger \hat a_-$, have frequencies
\begin{equation}\label{eq_omega_Z4_app}
    \omega_{\pm}=\frac{2e^2}{\sqrt{2}\,\pi C_g}\sqrt{|q|^{2}\pm\sqrt{|q|^{4}-4\big[\mathcal{C}^{(2)}\big]^{2}}}\ ,\qquad |q|^2=\sum_{z<z'}q_{zz'}^{2}\ .
\end{equation}
They are the values $(2e^2/\pi C_g)\lambda_\pm$ for the eigenvalues $\pm i\lambda_\pm$ of $\mathcal C^{(1)}$, and for block-diagonal $\mathcal C^{(1)}$ they reduce to two copies of Eq.~\eqref{eq_omega_Z2}.

\section{\bluehead{Coupling to external fields and gauge invariance}}\label{app:gauge_fields}

In the projected two-node Hamiltonian, Eq.~\eqref{eq_H_twonodes_external_fields}, an offset charge at one node and an offset flux at the other enter in the same combination. Offset charges and fluxes correspond to applied voltages and magnetic fields, described by potentials $\varphi$ and $\mathbf A$ that are defined only up to gauge transformations leaving $\mathbf E=-\nabla\varphi-\dot{\mathbf A}$ and $\mathbf B=\nabla\times\mathbf A$ invariant. Two features of Eq.~\eqref{eq_H_twonodes_external_fields} therefore require comment.

The first concerns static fluxes. The two junctions in Fig.~\ref{fig:spectrum}(a) appear to form a loop, which would suggest that only the loop phase $\phi_1^{(\mathrm{ext})}-\phi_2^{(\mathrm{ext})}$ matters. In Eq.~\eqref{eq_H_twonodes_external_fields}, both phases have separate, measurable effects. This is consistent with gauge invariance when the gyrating element is a separate device coupled to ground, as in Fig.~\ref{fig:qcgyrator}(c): the gyrator and the two junctions then form two distinct loops. If instead the Josephson coupling is intrinsic to the gyrating element, there is no loop and the flux dependence disappears. This case is described by shifting $\phi_z$ by $\phi_z^{(\mathrm{ext})}$ everywhere in the Lagrangian, not only in $\epsilon_0$. The flux response therefore distinguishes intrinsic from external Josephson elements.

The second concerns offset charges. Because $\mathbf E=-\nabla\varphi-\dot{\mathbf A}$, offset charges can usually be converted into time-dependent flux offsets by a unitary transformation~\cite{You2019,Riwar_2021}. This is not possible for Eq.~\eqref{eq_H_twonodes_external_fields}. The coupling of the effective low-energy theory to electromagnetic fields must therefore be derived from the circuit, as done here, rather than imposed by gauge arguments on the effective theory.

\section{\bluehead{Protocol to probe the no-communication argument for encoded fermions}}\label{app:protocol_nocommun}

As pointed out in the main text, Ref.~\cite{johansson2016comment} introduces a simple protocol which demonstrates how breaking Wigner superselection allows for changing the outcome of a measurement in one subsystem (Alice), by only performing local gates in another subsystem (Bob), acting on different nodes. We first reiterate the protocol as given in Ref.~\cite{johansson2016comment}, and then we show how it is implemented within the circuit hardware.

The protocol only requires two fermionic modes, $a,a^\dagger$ for Alice, and $b,b^\dagger$ for Bob, such that $\{a,a^\dagger\}_{+}=\{b,b^\dagger\}_{+}=1$, whereas $\{a,b^\dagger\}_{+}=0$ (the two modes belong to different nodes). Starting from the vacuum state $|0\rangle$ ($a|0\rangle=b|0\rangle=0$), we first prepare the initial state with a unitary gate performed by Alice, $U_A=(1-a+a^\dagger)/\sqrt{2}$. This gate is designed to explicitly break Wigner superselection, since the initial state after the gate reads,
\begin{equation}
    |\psi_A\rangle=U_A|0\rangle=\frac{|0\rangle+a^\dagger |0\rangle}{\sqrt{2}}\ .
\end{equation}
We now imagine that Alice performs a local measurement with the observable $O_A=a+a^\dagger$, which yields,
\begin{equation}
    \langle\psi_A|O_A|\psi_A\rangle=1\ .
\end{equation}
If, before Alice's measurement of $O_A$, we allow Bob to perform a gate of the form $U_B=b+b^\dagger$, yielding the state $|\psi_{AB}\rangle=U_BU_A|0\rangle$, the subsequent measurement on Alice's side would yield
\begin{equation}\label{eq_psi_A}
    \langle\psi_{AB}|O_A|\psi_{AB}\rangle=-\langle\psi_A|O_A|\psi_A\rangle=-1\ ,
\end{equation}
which is manifestly different from the first measurement, demonstrating that here, the local gate applied on Bob's side immediately influences Alice's measurement, which the no-communication argument for local fermionic modes excludes.

The above protocol can be significantly generalized to an arbitrary number of fermionic modes, and involving any operator measured on Alice's side $O_A$, and any gate $U_B$ applied on Bob's side, as long as both $O_A$ and $U_B$ are odd in fermion ladder operators. Note that while $O_A,U_B$ being odd means that they change the fermion parity when being applied to a prepared state $|\psi_A\rangle$, they themselves do not violate the parity \textit{superselection} rule: they change the parity of any state they are applied to, but they do not create superpositions of states with different parity. They do however always anticommute, such that we are guaranteed to find $\langle \psi_A|U_B^\dagger O_A U_B|\psi_A\rangle = -\langle \psi_A|O_A|\psi_A\rangle$. Hence, the only condition for this protocol to violate no-signalling is that $\langle \psi_A|O_A|\psi_A\rangle\neq 0$. Such a nonzero expectation value of an odd fermion operator can in turn only emerge if $|\psi_A\rangle$ is in an even-odd superposition.

We now briefly outline, how this protocol is implemented. For this purpose, we consider the Majorana circuit (i.e., all Chern numbers equal $2$) in the main text, with four nodes, $Z=4$. We assign odd nodes $1$ and $3$ to Alice, and even nodes $2$ and $4$ to Bob. Choosing the same representation in terms of regular complex fermions as introduced in the main text, we identify $a=c_1=(\gamma_1-i\gamma_3)/2$ and $b=c_2=(\gamma_2-i\gamma_4)/2$. The protocol starts from the encoded vacuum $|0\rangle$, defined by $a|0\rangle=b|0\rangle=0$. The gate for the state initialization according to Eq.~\eqref{eq_psi_A} is performed straightforwardly. We switch on a Josephson effect between ground and node 3 (on Alice's side), yielding the Hamiltonian
\begin{equation}
    H_{A}=-2\widetilde{E}_{J}\gamma_{3}=-2i\widetilde{E}_{J}\left(a-a^{\dagger}\right)\ .
\end{equation}
Applying this Hamiltonian for time $\tau=\pi/(8\widetilde{E}_{J})$ yields exactly the unitary time-evolution $U_A$ as defined above. A Josephson junction between node $1$ and ground couples to the projected observable $O_A=\gamma_1=a+a^\dagger$. We assume a projective readout that distinguishes its two eigenvalues. Bob can implement the gate $U_B$ in two equivalent ways. Either, he switches on a Josephson effect between ground and node 2,
\begin{equation}
H_{B}=-2\widetilde{E}_{J}\gamma_{2}=-2\widetilde{E}_{J}\left(b+b^{\dagger}\right)\ ,
\end{equation}
this time, for double the time duration $\tau=\pi/(4\widetilde{E}_{J})$, or alternatively, he offsets the charge at node 2, $n_{2}^{\left(\text{ext}\right)}$, by $\pm1$. As demonstrated above, the expectation value measured on Alice's side swaps sign, depending on whether or not Bob performs gate $U_B$. The sign flip thus tests the anticommutation of the encoded operators $\gamma_1$ and $\gamma_2$ directly. Alice and Bob control disjoint sets of circuit nodes, while their projected operators act on the same collective Landau-level manifold. The Chern couplers supply the anticommutation and play the role of the Jordan--Wigner string of a qubit encoding, see Sec.~\ref{sec_fermion_parity}.

\bibliography{references}

\end{document}